\documentclass[preprint,12pt]{elsarticle}

\usepackage{psfrag}
\usepackage[T1]{fontenc}
\usepackage{color}
\usepackage{times}
\usepackage{amsmath}
\usepackage{amssymb}
\usepackage{graphicx}
\usepackage{wasysym}
\usepackage{textcomp}

\begin{document}

\begin{frontmatter}

\title{Critical Loop Gases and the Worm Algorithm}

\author[itp]{Wolfhard Janke}
\author[nic]{Thomas Neuhaus}
\author[itp]{Adriaan M. J.~Schakel}
\address[itp]{Institut f\"ur Theoretische Physik, Universit\"at Leipzig,
Postfach 100 920,
D-04009 Leipzig, Germany}
\address[nic]{J\"ulich Supercomputing Centre, 
  Forschungszentrum J\"ulich, D-52425 J\"ulich, Germany}
\begin{abstract}
  The loop gas approach to lattice field theory provides an alternative,
  geometrical description in terms of fluctuating loops.  Statistical
  ensembles of random loops can be efficiently generated by Monte Carlo
  simulations using the worm update algorithm.  In this paper, concepts
  from percolation theory and the theory of self-avoiding random walks
  are used to describe estimators of physical observables that utilize
  the nature of the worm algorithm.  The fractal structure of the random
  loops as well as their scaling properties are studied.  To support
  this approach, the O(1) loop model, or high-temperature series
  expansion of the Ising model, is simulated on a honeycomb lattice, with
  its known exact results providing valuable benchmarks.
\end{abstract}


\end{frontmatter}

\section{Introduction}
Representing the hopping of particles from one lattice site to the next,
the strong-coupling expansion in relativistic quantum field theories
formulated on a spacetime lattice provides an alternative approach to
numerically simulating lattice field theories in terms of world lines.
The standard approach, which is rooted in the functional integral
approach to field quantization, involves estimating observables
(expressed in terms of the fields) by sampling a representative set of
field configurations.  New configurations are typically generated by
means of a Monte Carlo technique which uses importance sampling, with
each field configuration weighted according to the probability that it
occurs.  In contrast, the approach based on the strong-coupling, or
hopping expansion, which is closely connected to Feynman's spacetime
approach to quantum theory \cite{Feynman}, involves linelike objects.
Physical observables are in this geometrical approach no longer
estimated by sampling an ensemble of field configurations, but by
sampling a grand canonical ensemble of (mostly closed) world lines,
known as a \textit{loop gas}, instead.  The weight of a given world line
configuration is typically determined by the total length of the paths,
the number of intersections, and the number of loops contained in the
tangle.

In statistical physics, the strong-coupling expansion is known as the
high-temperature series expansion \cite{Stanley}.  Lattice field
theories studied in this context are typically spin models, such as the
O($N$) spin model, whose representation in terms of high-temperature
(HT) graphs is known as a \textit{loop model}.

\begin{figure}
\centering
\includegraphics[width=0.6\textwidth]{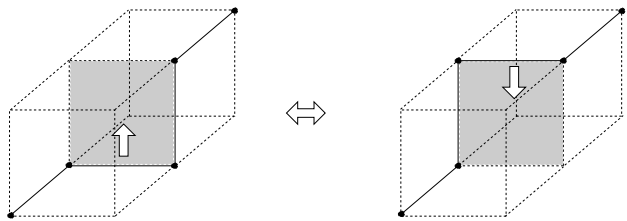}
\caption{Bond-shifting algorithm for generating a new world line
  configuration on a cubic lattice.  Lattice sites visited by the walk are marked by full circles and the updated plaquettes are shaded.
  \label{fig:shift}}
\end{figure}

A first numerical study of loop gases formulated on the lattice was
carried out by Berg and Foerster \cite{BB}.  New world line
configurations were generated by a bond-shifting Monte Carlo update
algorithm as follows.  A randomly chosen bond of the existing
configuration is shifted perpendicular to itself by one lattice spacing
in any of the $2(d-1)$ directions of the hypercubic lattice.  During the
shift, each of the endpoints of the moving link erases or draws a bond
in the chosen perpendicular direction, depending on whether the link is
occupied or not, as in Fig.~\ref{fig:shift}.  The new configuration is
accepted or rejected according to the Metropolis algorithm. 

At about the same time, Dasgupta and Halperin \cite{DaHa}, following a
suggestion by Helfrich and M\"uller \cite{HM} that the HT graphs of the
O($N$) lattice model simultaneously describe a loop gas of sterically
interacting physical lines, simulated a gas of directed loops on a cubic
lattice.  New loop configurations were generated in this study by
inserting an elementary loop, or \textit{plaquette}, of random
orientation according to the Metropolis algorithm.  

Although these and related early loop gas update algorithms
\cite{Karowski_etal,HJK,Elser} work fine in the disordered phase away from
the critical point, they all, being based on local updates, suffer from
pronounced critical slowing down.  That is, consecutive configurations
are highly correlated close to the critical point, and simulations on
larger lattices become increasingly unfeasible in this region.

About a decade ago, Prokof'ev and Svistunov~\cite{ProkofevSvistunov}
have introduced a Monte Carlo update algorithm that, although based on
local updates, does away with critical slowing down almost completely.
The so-called \textit{worm algorithm} generates loop configurations, not
by inserting plaquettes, but through the motion of the end points of an
\textit{open} world line---the ``head'' and ``tail'' of a ``worm''.  An
additional loop is generated in this scheme when the head bites the
tail, or through a ``back bite'' where the head erases
a piece (bond) of its own body and thereby leaves behind a detached loop
and a shortened open chain.

Besides this outstanding technical advantage, the worm algorithm has the
additional advantage in the context of statistical physics that the
complete set of standard critical exponents can be determined at a
stroke.  This set is known to split into two, \textit{viz.} the thermal
and the magnetic exponents.  While the thermal exponents, such as the
specific heat exponent $\alpha$, pertain to closed paths, the magnetic
exponents, such as the magnetic susceptibility exponent $\gamma$,
pertain to open paths in the geometrical approach.  Using a plaquette
update, one is restricted to the topology of the initial configuration.
If that starting configuration consists of just closed paths, a
plaquette update algorithm will subsequently also generate only loop
configurations.  Open paths, needed to determine the magnetic exponents,
must be sampled in such a scheme by putting in an open path connecting
two fixed endpoints from the start.  A plaquette update will then change
the loops fluctuating in the background and will also change the form of
the open path, but it will leave the endpoints of the path untouched.
Since, in principle, all possible end-to-end distances are needed to
determine the magnetic exponents, a plaquette update is impracticable to
achieve this.  By the nature of the worm algorithm, which features an
open path between loop updates, these data are generated on the fly in
this scheme.  More specifically, the open paths directly sample the
spin-spin, or two-point, correlation function.

In this paper, which extends previous work by two of us on the subject
\cite{geoPotts,ht}, we describe estimators of physical observables that
naturally arise in a loop gas and that allow determining the standard
critical exponents.  Our approach, put forward in Sec.~\ref{sec:l_n},
amalgamates concepts from percolation theory---the paradigm of a
geometrical phase transition---and the theory of self-avoiding random
walks.  We relate this geometrical approach to phase transitions in
terms of fluctuating paths to the more familiar field theory approach by
considering the O($N$) symmetric $\phi^4$ theory in Sec.~\ref{sec:phi4}.
To support our arguments, the second part of the paper is devoted to
Monte Carlo simulations of the two-dimensional O(1) loop model using the
worm update algorithm.  This model serves as a prototype with its
various exact results providing a yardstick for our Monte Carlo results
and also for the feasibility of our approach.  Section~\ref{sec:details}
specifies the model we have simulated, introduces the specific
implementation of the worm update algorithm used, and gives details of the
simulations.  Our results are presented in Sec.~\ref{sec:results}.  We
finish with a discussion and outlook.

\section{Loop gases}
\label{sec:l_n}
We are concerned with lattice field theories close to the critical point
where they undergo a continuous phase transition.  Their equivalent loop
gas representation can be conveniently characterized by the average
number $\ell_n$ of closed paths, or polygons, of $n$ steps per unit
volume.  Close to the critical point $K_\mathrm{c}$, the so-called loop,
or loop length distribution takes asymptotically a form \cite{geoPotts}
similar to the cluster distribution near the percolation threshold known
from percolation theory \cite{StauferAharony},
\begin{equation}
  \label{elln}
  \ell_n \sim n^{-d/D -1} \mathrm{e}^{- \theta n}, \quad \theta \propto
  (K-K_\mathrm{c})^{1/\sigma}.
\end{equation} 
Here, $\theta$ is the line tension (in suitable units), $K$ is the
tuning parameter, and $d$ denotes the dimension of space (in the case of
classical theories) or spacetime (in the case of quantum theories). When
the line tension is finite, the Boltzmann factor in the distribution
(\ref{elln}) exponentially suppresses long loops.  Upon approaching the
critical point, $\theta$ vanishes at a rate determined by the exponent
$\sigma$.  At $K_\mathrm{c}$, loops proliferate for they can now grow
without energy penalty.  The remaining factor in the loop distribution
is an entropy factor, giving a measure of the number of ways a polygon
of $n$ steps can be embedded in the lattice.  It is characterized by the
fractal dimension $D$ of the paths at the critical point.  The entropy
factor decreases with increasing $n$.

A standard definition of the fractal dimension is through the asymptotic
behavior of the average square radius of gyration $\langle
R_\mathrm{g}^2\rangle$ of chains of $n$ steps as
\begin{equation} 
\label{rms}
\langle R_\mathrm{g}^2\rangle  \sim n^{2/D},
\end{equation}   
where 
\begin{equation}
\label{Rg} 
  R_\mathrm{g}^2  \equiv \frac{1}{2n^2} \sum_{k,k'=1}^n  (x_{i_k} -
    x_{i_{k'}})^2  = \frac{1}{n} \sum_{k=1}^n  (x_{i_k} -
    \bar{x})^2 
\end{equation} 
with $x_{i_k}$ the position vector of the chain after $k$ steps and
\begin{equation}
\label{center} 
\bar{x} \equiv \frac{1}{n} \sum_{k=1}^n  x_{i_k}
\end{equation} 
the center of mass of the chain (which can be closed or open) of $n$
steps.  Here and in the following, lattice sites are labeled by the
index $i$.  The radius of gyration gives a measure of the distance
covered by the path.  Another standard definition is through the average
square end-to-end distance $\langle R^2_\mathrm{e} \rangle$ of open
chains of $n$ steps,
\begin{equation}
\label{e2e} 
  \langle R_\mathrm{e}^2\rangle \equiv \left\langle (x_{i_n} -
  x_{i_0})^2 \right\rangle \sim n^{2/D},
\end{equation} 
where $x_{i_0}$ denotes the starting point of the chain.  For the
two-dimensional O($N$) model, which for $-2 \leq N \leq 2$ undergoes a
continuous phase transition, the fractal dimension $D$ of the HT graphs
\cite{Vanderzande} corresponds to the renormalization group eigenvalue
$y_2$ of the two-leg operator in the spin representation of the model
\cite{Nienhuis_rev}.

The natural length scale in quantum field theory is the correlation
length $\xi$.  The critical exponent $\nu$, characterizing the
divergence of this length scale when the critical point is approached,
$\xi \sim |K-K_\mathrm{c}|^{-\nu}$, is related to the fractal dimension
through \cite{ht}
\begin{equation}
  \label{nuD}
  \nu = 1/\sigma D .
\end{equation}
This expression, which assumes the same form as in percolation theory
\cite{StauferAharony}, generalizes a celebrated result due to de~Gennes
\cite{deGennes} for self-avoiding random walks (SAWs), which corresponds
to the limit $N\to0$ of the O($N$) spin model.  In that case, $\sigma
=1$, but in general $\sigma$ takes different values, see
Table~\ref{table:On} below.

As is known from the theory of SAWs, closed paths alone yield only the
thermal exponents of the universality class defined by the O($N\to 0$)
model.  To obtain also the magnetic exponents, and thereby the
complete set of standard exponents, the total number
\begin{equation} 
\label{zn_def}
  z_n \equiv
  \sum_{j} z_n(x_i, x_j)
\end{equation}  
of SAWs of $n$ steps starting at $x_i$ and ending at an arbitrary site
$x_j$ is needed in addition.  Because of translational symmetry, $z_n$
does not depend on $x_i$, and $z_n(x_i, x_j)$ only depends (up to
lattice artifacts) on the end-to-end distance $r \equiv |x_i - x_j|$,
i.e., $z_n(x_i, x_j) = z_n(r)$.  The ratio of $z_n(x_i, x_j)$ and $z_n$
defines the probability $P_n(x_i, x_j)$ of finding a chain connecting
$x_i$ and $x_j$ in $n$ steps.  As for SAWs \cite{Fisher}, we expect this
distribution to scale for a general loop gas as
\begin{equation}
  \label{P}
  P_n(x_i, x_j) \equiv z_n(x_i, x_j)/z_n
  \sim n^{-d/D} \, \mathcal{P} \bigl( r/n^{1/D}
  \bigr),
\end{equation} 
with $\mathcal{P}$ a scaling function.  That is, we assume that
$P_n(x_i, x_j)$ depends only on the ratio $r/\langle
R_\mathrm{g}^2\rangle^{1/2}$.  As an aside, the average square end-to-end
distance (\ref{e2e}) is the second moment of this distribution.  In
continuum notation:
\begin{equation} 
  \langle R_\mathrm{e}^2\rangle = \Omega_d \int_0^\infty \mathrm{d} r \, r^{d-1} \, r^2 P_n(r) ,
\end{equation} 
where $\Omega_d$ denotes the surface of a unit hypersphere embedded in
$d$ space dimensions, and $P(r)$, being a probability, is normalized to
unity
\begin{equation} 
\label{norm}
1 = \Omega_d \int_0^\infty \mathrm{d} r \, r^{d-1}  P_n(r) .
\end{equation} 

In addition to the scaling (\ref{P}), we also assume the number $z_n$ to
scale as
\begin{equation}
\label{zn}
  z_n K^n \sim n^{\vartheta/D} {\rm e}^{- \theta n}
\end{equation}   
with a universal exponent $\vartheta$ that characterizes, as do the rest
of the critical exponents, the universality class.  For the O($N$)
model, it depends, in addition to the dimensionality $d$, solely on $N$.
Since the number of possible rooted open chains with no constraint on
their endpoint increases with the number $n$ of steps, $\vartheta$ is
expected to be positive. This is in contrast to closed chains, where the
corresponding factor in Eq.~(\ref{elln}) decreases with increasing $n$,
reflecting that it becomes increasingly more difficult for chains to
close the longer they are.

The fractal dimension $D$ together with the exponents $\sigma$ and
$\vartheta$ determine the standard critical exponents of the theory.  As
for SAWs, the relevant scaling relations can be derived by writing the
correlation function $G(x_i, x_j)$ as a sum over all possible chains of
arbitrary many steps joining the endpoints:
\begin{equation} 
\label{Gsum}
G(x_i, x_j) = \sum_n z_n (x_i, x_j) K^n .
\end{equation} 
As before, $G(x_i, x_j) = G(r)$ because of translational invariance.
When evaluated at the critical point, where the correlation function
depends algebraically on the end-to-end distance, $G(x_i, x_j) \sim
1/r^{d-2 + \eta}$, this gives
\begin{equation} 
\label{vartheta}
\eta = 2 - D - \vartheta.
\end{equation}
Given the exact values for $\eta$ \cite{Nienhuis} and the fractal
dimension $D$ of the HT graphs \cite{Vanderzande}, $\vartheta$ can be
determined exactly for the two-dimensional O($N$) model, see
Table~\ref{table:On}.  Through the exact enumeration and analysis of the
number $z_n$ of SAWs on a square lattice up to length 71, the expected
value $\vartheta/D=\frac{11}{32}$ for $N=0$ has been established to high
precision \cite{Jensen04}.

\begin{table}
  \caption{Critical exponents of the two-dimensional critical O($N$)
    spin models, with $N=-2,-1,0,1,2,\infty$, respectively, together with the 
    fractal dimension $D$ of the HT graphs as well as the two exponents
    $\sigma$ and $\vartheta$.
    \label{table:On}}
\centering
  \begin{tabular}{l|r|ccc|ccc}
%
%    \hline \hline & & & & & & \\[-.4cm]
%
    Model & $N$ & $\gamma$ & $\eta$ & $\nu$ & $D$ & $\sigma$ &
    $\vartheta$ \\[.05cm]
    \hline & & & & & & \\[-.4cm]
    Gaussian & $-2$ & $1$ & $0$ & $\frac{1}{2}$ & $\frac{5}{4}$ &
    $\frac{8}{5}$ & $\frac{3}{4}$\\[.1cm]
             & $-1$ & $\frac{37}{32}$ & $\frac{3}{20}$ & $\frac{5}{8}$ & $\frac{13}{10}$ &
    $\frac{16}{13}$ & $\frac{11}{20}$\\[.1cm]
    SAW & $0$ & $\frac{43}{32}$ & $\frac{5}{24}$ & $\frac{3}{4}$ &
    $\frac{4}{3}$ & $1$ & $\frac{11}{24}$ \\[.1cm]
    Ising & $1$ & $\frac{7}{4}$ & $\frac{1}{4}$ & $1$ & $\frac{11}{8}$ &
    $\frac{8}{11}$ & $\frac{3}{8}$ \\[.1cm]
    XY & $2$ & $\infty$ & $\frac{1}{4}$ & $\infty$ & $\frac{3}{2}$ & $0$
    & $\frac{1}{4}$ \\[.1cm]

    Spherical & $\infty$ & $\infty$ & $0$ & $\infty$ & $2$ & $0$
    & $0$ 
% \\[.1cm] \hline \hline
% 
%
  \end{tabular}
\end{table}

The relation (\ref{nuD}) can, incidentally, be derived by using the
second-moment definition of the correlation length $\xi$,
\begin{equation} 
\label{xi2}
\xi^2 = \frac{\int_0^\infty \mathrm{d} r \, r^{d-1} \, r^2 
G(r)}{\int_0^\infty \mathrm{d} r \, r^{d-1} \, G(r)}
\end{equation} 
in continuum notation.
Finally, using the definition of the susceptibility $\chi$, $\chi =
\sum_{j} G(x_i, x_j)$, which diverges as $\chi
\sim |K-K_\mathrm{c}|^{-\gamma}$, we find
\begin{equation} 
\label{gamma}
\gamma = (D + \vartheta)/\sigma D .
\end{equation} 
This relation generalizes one originally due to des~Cloizeaux
\cite{Cloizeaux} for SAWs for which $\sigma=1$.  The explicit
expressions for $\nu,\eta$, and $\gamma$ satisfy Fisher's scaling
relation, $\gamma/\nu = 2-\eta$.  Note that only the combinations
$D+\vartheta$ and $\sigma D$ enter the scaling relations between the
various critical exponents.

\begin{figure}
\centering
\psfrag{x}[t][t][1][0]{$x$}
\includegraphics[width=0.3\textwidth]{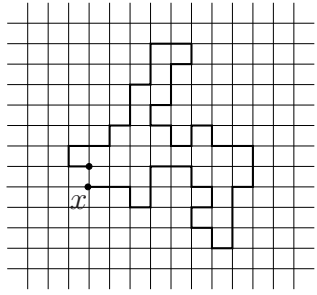}
\caption{A SAW on a square lattice returning to a site adjacent to its
  starting point $x$.
  \label{fig:saw}}
\end{figure}

We next consider the limit $x_j \to x_i$ of $z_n (x_i, x_j)$.  Following
standard practice in the theory of SAWs \cite{deGennesbook}, we define
this limit of vanishing end-to-end distance as the number of chains
$z_n(x_i,x_i \pm a \hat{\mu})=z_n(a)$ of $n$ steps returning to a site
$x_i \pm a \hat{\mu}$ adjacent to the starting point $x_i$.  Here, $\pm
\hat{\mu}$ is a unit vector in any (positive as well as negative)
direction on the lattice (see Fig.~\ref{fig:saw}), and $a$ is the
lattice spacing.  That is, $z_n(a)$ rather than $z_n(0)$ is taken when
closing open chains, with the lattice spacing $a$ serving as an
ultraviolet cutoff.  In continuum quantum field theory, the limit $x'
\to x$ corresponds to putting two fields at the same point.  Such
composite operators usually need special care and require a separate
renormalization independent of that of the constituting operators.  The
number of chains $z_n(a)$ is related to the loop distribution
(\ref{elln}) through
\begin{equation}
\label{ellz}
\ell_n = \frac{1}{n} z_n(a) K^n .
\end{equation} 
Since a polygon can be traced out starting at any lattice site along the
chain, the factor $1/n$ is included to avoid double counting.  Note that
the loop distribution is a density being defined per lattice site and
that $z_n(a)$ refers to \textit{rooted} closed chains all starting at
the same lattice site $x_i$.  As first shown by McKenzie and Moore
\cite{McKenzieMoore} for SAWs, consistency of Eq.~(\ref{ellz}) with
$z_n(a) = z_n P_n(a)$ and Eq.~(\ref{elln}) requires that the scaling
function $\mathcal{P}(t)$ must vanish for $t \to 0$ and behave for
\textit{small} argument $t$ as
\begin{equation} 
\label{small}
\mathcal{P}(t) \sim t^\vartheta
\end{equation}  
with an exponent determined by the \textit{asymptotic} behavior
(\ref{zn}) of the number $z_n$ of open chains at the critical point.
With this identification, Eq.~(\ref{vartheta}) becomes the relation
first proposed by Prokof'ev and Svistunov in
Ref.~\cite{ProkofevSvistunov_com}.  Together with the relation
(\ref{nuD}) proposed in Ref.~\cite{ht}, Eq.~(\ref{vartheta}) allows 
expressing the standard critical exponents in terms of the fractal
structure of open and closed paths and the rate $1/\sigma$ at which the
line tension vanishes upon approaching the critical point.  As already
mentioned in the Introduction, a major advantage of the worm update
algorithm is that it features both open and closed paths because this makes
possible to determine all these exponents at a stroke.

\section{$|\phi^4|$ Lattice field theory}
\label{sec:phi4}
To make connection with field theory, we consider as an example the
O($N$) symmetric $\phi^4$ theory formulated on a hypercubic lattice in
$d$ Euclidean spacetime dimensions.  The theory is specified by the
(Euclidean) lattice action
\begin{equation} 
\label{Sor}
S = a^d \sum_i \left\{ \frac{1}{2a^2} \sum_\mu \left[\varphi(x_i+a
  \hat{\mu}) - \varphi(x_i)\right]^2 + \frac{m^2}{2} \varphi^2(x_i) +
  \frac{g}{4!}  \varphi^4(x_i) \right\}
\end{equation} 
with lattice spacing $a$.  The real scalar field $\varphi(x_i)$, which
is defined on the lattice sites $x_i$ of the spacetime box, has $N$
components $\varphi = \varphi^\alpha = (\varphi^1, \varphi^2, \ldots ,
\varphi^N)$.  As before, the index $i$ labels the lattice sites, the sum
$\sum_i$ stands for a sum over all lattice sites, and the index
$\alpha=1,2, \ldots, N$ labels the field components.  Moreover,
$\varphi^4 \equiv (\varphi \cdot \varphi)^2$, where the dot product
implies a summation over the field components: $\varphi \cdot \varphi =
\sum_{\alpha=1}^N \varphi^\alpha \varphi^\alpha$.  Lattice coordinates,
representing discretized spacetime, are specified by $x_i = x_i^\mu=
(x^1, x^2, \ldots , x^d)_i$, with $\hat{\mu}$ denoting the unit vector
pointing in the (positive) $\mu$-direction.  Moreover, $m^2$ is the bare
mass parameter squared, and $g$ is the bare coupling constant of the
self-interaction term.  In the world line picture, this four-leg
operator corresponds to intersections where two lines cross.  The
renormalization group eigenvalue $y_4$ of the four-leg operator
corresponds to the fractal dimension $D_\times$ of these intersections.
Numerically, this fractal dimension can be determined through
finite-size scaling by measuring the average number, or ``mass''
$M_\times$, of these intersections which scales at the critical point as
\begin{equation} 
M_\times(L) \sim L^{D_\times} 
\end{equation} 
with the linear size $L$ of the lattice.  For the critical
two-dimensional O($N$) model, the four-leg operator is irrelevant for
$-2 \leq N < 2$, i.e., $y_4 = D_\times<0$, and it becomes marginal for
$N=2$ \cite{Nienhuis_rev}.  Because intersections are irrelevant (or
marginal) there, loops at the O($N$) critical point are frequently
referred to as \textit{dilute} loops.

In the continuum limit, where the lattice spacing tends to zero, $a
\to 0$, the lattice action (\ref{Sor}) reduces to the standard form
\begin{equation} 
\label{Scont}
S = \int \mathrm{d}^d x \left\{ \frac{1}{2} \left[\partial_\mu \varphi(x)
\right]^2 + \frac{m^2}{2} \varphi^2(x) + \frac{g}{4!}  \varphi^4(x)
\right\},
\end{equation}
where $\varphi(x)$ stands for the field defined in continuous spacetime.

The partition function $Z$ of the lattice theory obtains by carrying
out the sum, or integral over the spin variable at each site of the
lattice:
\begin{equation} 
\label{Z}
Z = \mathrm{Tr} \, \mathrm{e}^{-S},
\end{equation} 
with
\begin{equation} 
\label{Tr}
  \mathrm{Tr} \equiv \prod_i \int \mathrm{d}^N \varphi(x_i).
\end{equation}   
This amounts to summing, or integrating over all possible spin
configurations, each weighted by the Boltzmann factor $\mathrm{e}^{-S}$
(in natural units).  In the continuum limit $a \to 0$, this defines the
functional measure $\int\mathrm{D} \varphi$, and the partition function
becomes
\begin{equation} 
\label{partition}
Z = \int \mathrm{D} \varphi \, \mathrm{e}^{-S} .
\end{equation}

For numerical simulations, a more convenient form of the lattice action
is obtained by casting Eq.~(\ref{Sor}) in terms of dimensionless fields
and parameters defined through \cite{JanSmit}
\begin{eqnarray} 
\label{K1}
a^{d-2} \varphi^2(x_i) &=& 2 K \, \phi_i^2 \\ 
\label{K2} 
a^{4-d} g &=& 6 \frac{\lambda}{K^2} \\ 
\label{K3}
m^2 a^2  &=& \frac{1-2\lambda N}{K} -2 d ,
\end{eqnarray}  
with $K>0$.  The action then takes the form of an O($N$) spin model
\begin{equation}
\label{Sorp}
S = - K \sum_{\langle i,i' \rangle} \phi_i \cdot
  \phi_{i'} + \sum_i \phi^2_i + \lambda \sum_i \left(
  \phi_i^2 - N \right)^2 .
\end{equation}
The sum $\sum_{\langle i,i' \rangle}$ extends over all nearest neighbor
pairs.  In terms of these new dimensionless variables, the action is
independent of the lattice spacing $a$.  The partition function $Z$ can
now be written as
\begin{equation} 
\label{full}
  Z = \int \mathrm{D} \mu(\phi) \exp \Biggl(K \sum_{\langle i,i'
    \rangle} \phi_i \cdot \phi_{i'} \Biggr),
\end{equation} 
with the on-site measure
\begin{equation}
\label{onsite} 
  \int \mathrm{D} \mu(\phi) \equiv \int \prod_i \mathrm{d}^N \phi_i \;
      \mathrm{e}^{-\phi^2_i - \lambda \left( \phi_i^2 - N \right)^2}.
\end{equation} 
In the limit $\lambda \to \infty$, the lattice field theory reduces to
the standard O($N$) spin model\index{O($N$) model}, with a ``spin''
variable $\phi_i$ of fixed length, $\phi_i^2 = N$, located at each site
of the spacetime lattice.  The remaining factor in the on-site measure
(\ref{onsite}) becomes trivial in this limit and can be ignored.
The normalization is chosen such that
\begin{equation} 
  \int \mathrm{d}^N \phi_i = 1, \quad \int \mathrm{d}^N \phi_i \, \phi_i^2 = N.
\end{equation}

Instead of considering the conventional Boltzmann weight factor, often a
simplified representative of the O($N$) universality class is studied,
obtained by truncating that factor \cite{DMNS}:
\begin{equation}
  \label{ZS}
  Z = \int \prod_i \mathrm{d}^N \phi_i \prod_{\langle i,i' \rangle} (1
  + K \phi_i \cdot \phi_{i'}) .
\end{equation} 
The second product is restricted to nearest neighbor pairs.  The main
difference with the original spin model is that in the truncated model,
links cannot be multiply occupied.  The weight carried by a
configuration is positive for $|K|<1/|N|$.  By universality, the truncated
model is expected to still belong to the O($N$) universality class.
Note that for $N=1$, where $\phi_i = \pm 1$, the full Boltzmann factor
can be exactly written in the truncated form by the identity
\begin{equation} 
  \mathrm{e}^{\beta \phi_i \phi_{i'}} = \cosh (\beta) \left[1 + \tanh (\beta)
  \phi_i \phi_{i'} \right] \propto 1 + K \phi_i \phi_{i'}
\end{equation} 
with $K = \tanh (\beta)$.  The prefactor $\cosh (\beta)$ is immaterial
and can be ignored as far as critical phenomena are concerned.  The worm
algorithm~\cite{ProkofevSvistunov} was originally designed to simulate
the HT representation of the theory (\ref{full}) with the
full Boltzmann factor included so that links can be multiply occupied.
However, as already suggested by its inventors~\cite{ProkofevSvistunov},
the algorithm can be readily adapted to simulate the truncated model
(\ref{ZS}) without multiple occupied links.

The scaling part of the logarithm of $Z$ reads expressed in terms of the
loop distribution
\begin{equation} 
\label{Zell}
\ln Z/V \sim \sum_n \ell_n,
\end{equation} 
with $V$ the volume.  The result (\ref{nuD}) immediately follows from
the hyperscaling argument that $\ln Z/V \sim \xi^{-d}$.  

In (continuum) quantum field theory, the two-point correlation function
$G(x,x')$ is given in the symmetric phase by the average of a product of
two $\varphi$ fields at $x$ and $x'$, respectively:
\begin{equation}
   G(x,x') \equiv \langle \varphi (x) \cdot  \varphi(x') \rangle .
\end{equation} 
Its algebraic behavior at the critical point is in this context
parameterized as $G(x,x') \sim 1/r^{2 d_\varphi}$ with
\begin{equation}
\label{dphi} 
d_\varphi = \tfrac{1}{2} ( d -2 + \eta) = \tfrac{1}{2} ( d - D - \vartheta)  
\end{equation} 
denoting the anomalous scaling dimension of the $\varphi$ field.  The limit
$x' \to x$ of the correlation function $G(x,x')$ is conventionally
defined through a ``mass insertion'' as \cite{Cloizeaux}
\begin{equation} 
\label{G0}
  G(0) = \langle \varphi^2(x) \rangle \propto
  - \frac{\partial}{\partial m^2} \ln Z .
\end{equation} 
By Eqs.~(\ref{Zell}) and (\ref{K3}) it then follows that for a loop gas
\begin{equation} 
\label{phi2}
  \langle \varphi^2(x) \rangle \sim  (K_\mathrm{c}-K)^{1/\sigma-1}
  \sum_n n \, \ell_n,
\end{equation} 
or
\begin{equation} 
  \langle \varphi^2(x) \rangle \sim  1/\xi^{d_{\varphi^2}},
\end{equation} 
with 
\begin{equation} 
\label{dphi2}
  d_{\varphi^2} = d - \frac{1}{\nu} 
\end{equation} 
the standard expression for the scaling dimension of the composite
operator $\varphi^2(x)$. In deriving this, use is made of the relation
(\ref{nuD}).  

Note that naively taking the limit $x' \to x$ in Eq.~(\ref{Gsum})
yields, after using Eq.~(\ref{ellz}), the result (\ref{phi2}) without
prefactor.  This is the world line counterpart of the observation that
composite operators usually require a multiplicative renormalization by
themselves that cannot be expressed in terms of the renormalization
factors of the constituting operators.  Also the power-law decay
(\ref{small}) of the scaling function $\mathcal{P}(t)$ for $t \to 0$ is
related to this.  Assuming that the scaling function remains finite in
the limit $t \to 0$, one obtains from Eq.~(\ref{Gsum}) with the relation
(\ref{P}) the incorrect result
\begin{equation} 
  G(0) \sim \sum_n z_n P_n(a) K^n \sim 1/\xi^{2 d_\varphi} , \qquad
  \mathrm{(incorrect)}
\end{equation} 
involving the anomalous dimension of $\varphi$ instead of $\varphi^2$.  Only
for noninteracting theories, the renormalization of composite operators
can be expressed in terms of the renormalization of the constituting
operators, and $\mathcal{P}(0)$ is nonzero.

Noting that the right side of Eq.~(\ref{G0}) physically denotes the
internal energy, we conclude from Eq.~(\ref{phi2}) that in the world
line approach this quantity is determined by the average number of bonds
in closed graph configurations, i.e., by the average total loop length
in configurations without open chains.

In closing this section, we remark that the two combinations
$D+\vartheta$ and $\sigma D$, on which the standard critical exponents
depend, determine the anomalous scaling dimension of the $\varphi$ and
$\varphi^2$ fields through Eqs.~(\ref{dphi}) and (\ref{dphi2}) with
$1/\nu = \sigma D$, respectively.

\section{Model and details of simulation}
\label{sec:details}

\subsection{Loop model}
To specify the model we have simulated, we start with the 
representation (\ref{ZS}) of the O($N$) model.  Expanding the product
appearing there, one readily verifies that only terms with an
\textit{even} number of spins at each lattice site contribute to the
partition function.  A factor $K \phi^\alpha_i \phi^\alpha_j$ (no
summation over $\alpha$) in such a term can be conveniently visualized
by drawing a bond along the link of the underlying lattice connecting
the nearest neighbor sites labeled by $i$ and $j$.  With each field, or spin,
component $\alpha=1, 2, \ldots, N$ is associated a color, so that the
bonds come in $N$ colors.  Terms contributing to $Z$ then correspond to
closed graphs made up of such bonds and of vertices connecting an even
number of bonds.  The partition function is obtained by adding all these
contributions, i.e., by summing over all possible disconnected  closed graph
configurations, each carrying a certain weight.

The spin-spin, or two-point, correlation function $G(x_i,x_j)$ of the
truncated model
\begin{equation} 
\label{G1}
  G (x_i,x_j) = \left\langle \phi_{i} \cdot \phi_{j}
  \right\rangle = \frac{1}{Z} \int \prod_{i'} \mathrm{d} \phi_{i'}
  \, \phi_i \cdot \phi_j \prod_{\langle i', j' \rangle} (1 +
  K \phi_{i'} \cdot \phi_{j'})
\end{equation} 
can be treated in a similar fashion as the partition function with the
proviso that for terms in the expansion of the product in the numerator
to contribute, the two sites labeled by $i$ and $j$ must house, in
contrast to all other lattice sites, an odd number of spins.
Graphically, such terms typically correspond to a set of disconnected
closed graphs with an additional open graph connecting the two endpoints
$x_i$ and $x_j$.

The O($N$) loop model is obtained by resolving each closed graph into a
unique\-ly defined set of possibly intersecting loops.  This is done by
providing instructions how vertices connecting more than two bonds are
to be resolved.  In principle, such ``walking instructions'' can be
formulated on an arbitrary lattice in arbitrary dimensions \cite{CPS}.
However, the simplest way to deal with this issue is to consider a
honeycomb lattice, which has coordination number $z=3$, so that closed
graphs simply cannot intersect.  A configuration $\mathcal{G}$ of
disconnected closed graphs then automatically decomposes into loops, and
the partition function of the resulting loop gas assumes the form
\cite{DMNS}
\begin{equation}
\label{Zloop}
  Z_\mathrm{loop}= \sum_\mathcal{G} K^b N^l  ,
\end{equation}
where $b$ denotes the number of bonds and $l$ the number of loops in the
graph.  Each bond in a graph configuration carries a weight $K$, while
each loop carries a degeneracy factor $N$, for they can have any of the
$N$ colors.  These factors play the role of bond and loop fugacities in
the loop model.  The number of bonds in a graph configuration increases
with increasing bond fugacity $K$ and \textit{vice versa}.  The critical
point of the O($N$) loop model on a honeycomb lattice is exactly known
to be given by \cite{Nienhuis} 
\begin{equation}
\label{Kc}
K_\mathrm{c} = \left[2 + (2 -N)^{1/2}\right]^{-1/2} .
\end{equation} 

\subsection{Update algorithm}
In the main simulations, we restricted ourselves to the Ising model
($N=1$) on a honeycomb lattice.  From a loop gas perspective, this model
defines a statistical ensemble of polygons built from bond variables
$b_l$ which are defined on the links of the lattice.  Reflecting the
fact that only one color is present ($N=1$), the bond variables only
take the values $b_l=1$, when the bond labeled by $l$ is set, or
$b_l=0$, when it is not.  Apart from considering loop configurations, we
also consider configurations that have in addition a single chain
connecting two endpoints, $x_i$ and $x_j$ say.  Such configurations,
which correspond to two spin insertions in the spin representation,
contribute to the numerator $Z(x_i,x_j)$ of the spin-spin correlation
function
\begin{equation}
\label{G2} 
  G (x_i,x_j) =   \frac{Z (x_i,x_j)}{Z} 
\end{equation} 
and are naturally generated by the worm algorithm that locally updates
the bond configurations.  Although polygons on a honeycomb lattice
cannot intersect, an open chain can ``back bite'' or touch a polygon.
Such configurations, where a chain endpoint connects three bonds, are
allowed and must be included in the update scheme.  For a general loop
model, such configurations pose a problem, for they can lead to a change
in the number of loops during the next bond update.  Then to keep track
of the number of loops, the open chain must be traced out anew, making
the update algorithm nonlocal and slowing it down considerably.  The
Ising model is special in that the loop fugacity is unity, so that
configurations with lattice sites housing three bonds do not pose a
problem, at least not when just updating and not measuring them (see
below).
For the O(1) loop model, the updates involve
Metropolis flips of single bonds where the value $b_l$ of the bond
variable is replaced with $1-b_l$.  During the Monte Carlo simulation,
chain endpoints move and, thus, accumulate information about open chain
properties, such as their end-to-end distance.  As there is a finite
probability for an open chain to close and form a polygon, the algorithm
also acquires information about the loop gas.  We adapted the original
worm algorithm~\cite{ProkofevSvistunov} as follows, see
Refs.~\cite{Gabriel,Wolff} for related adaptations.

For configurations containing, in addition to polygons, a single chain
with end-to-end distance larger than one lattice spacing, the updating
scheme proceeds by
\begin{enumerate}
\item randomly choosing either endpoint of the chain,
\item randomly choosing any of the links attached to the chosen
  endpoint,
\item updating the corresponding bond variable $b_l$ with a single-hit
  Metropolis flip proposal $b_l \to b_l' = 1 - b_l$ with acceptance
  probability
\begin{equation}
\label{metropolis_acceptance}
P_\mathrm{accept}= \mathrm{min} \left(1,K^{1 - 2b_l} \right)
\end{equation}
as can be inferred from the weight $K^b$ in the partition function
(\ref{Zloop}), assuming that $0 < K < 1$.  The exponent $1-2b_l= \pm 1$
denotes the difference in the number of bonds contained in the proposed
and the existing configurations.  It follows that a proposal to
\textit{create} a bond is accepted with probability $P_\mathrm{accept}=
K \, (<1)$, whereas a proposal to \textit{delete} one is always
accepted.
\end{enumerate}
These updates are simple and straightforward as long as the chain
remains open.  Once, however, the chain has an end-to-end distance of
just one lattice spacing, the existing configuration can be turned into
a loop gas configuration by a single bond flip.  Such an update then
connects two different sectors of the model.  Namely, the sector with an
open path, which samples the numerator $Z(x_i,x_j)$ of the correlation
function (\ref{G2}), and the loop sector, which samples the partition
function $Z$.  In their original work~\cite{ProkofevSvistunov},
Prokof'ev and Svistunov introduced conditional probabilities,
parameterized by $0 < p_0 < 1$, for Monte Carlo moves between the two
sectors.  We in this work put this parameter to unity and, thus, always
attempt to close such a chain by using the update scheme above with the
Metropolis acceptance probability (\ref{metropolis_acceptance}).  If the
update is accepted, and the open chain turns into a polygon, we proceed
by randomly choosing one link among all links of the lattice.  The bond
variable on that link is then subjected to a Metropolis trial move with
the acceptance probability (\ref{metropolis_acceptance}).

To check the correctness of this worm algorithm, we simulated the
critical O(1) loop model on a small $5 \times 5$ square lattice with
periodic boundary conditions, i.e., on a torus and measured the
spin-spin correlation function.  Table~\ref{table:gii} summarizes our
Monte Carlo results and compares them to the exact results, obtained by
direct enumeration.  The table shows complete agreement within
statistical error bars.
\begin{table*}
\caption{Precision check of our worm update algorithm for the critical
  Ising model on a square $5 \times 5$ lattice with periodic boundary
  conditions.  We simulated the two-point correlation function
  $G(x_1,x_i)$ by putting one spin at the origin, labeled by $1$, and
  the other spin at the site labeled by the index $i$, which goes
  through the lattice in a typewriter fashion. The second column gives
  results from exact summations, while the third column summarizes our
  Monte Carlo data obtained with the worm algorithm. The fourth column
  shows our Monte Carlo data in units of the exact results, and the last
  column gives the bit-variable $t$ that is unity if theory and
  simulation differ by less than one $\sigma$, and is zero otherwise.}
\centerline{
\begin{tabular}{l|llll}
\hline
$i$ & $G_\mathrm{exact}(x_1,x_i)$ & $G(x_1,x_i)$ & $G/G_\mathrm{exact}$ &
$t$    \\
\hline
   1 &   1.000000 &  1.0           &  1.0           &   1      \\
   2 &   0.768360 &  0.768353(30)  &  0.999991(39)  &   1      \\
   3 &   0.708394 &  0.708385(23)  &  0.999987(33)  &   1      \\
   4 &   0.708394 &  0.708342(38)  &  0.999927(54)  &   0      \\
   5 &   0.768360 &  0.768354(32)  &  0.999993(41)  &   1      \\
   6 &   0.768360 &  0.768350(40)  &  0.999987(52)  &   1      \\
   7 &   0.722100 &  0.722082(38)  &  0.999976(52)  &   1      \\
   8 &   0.695433 &  0.695370(33)  &  0.999910(48)  &   0      \\
   9 &   0.695433 &  0.695422(39)  &  0.999986(56)  &   1      \\
  10 &   0.722100 &  0.722175(31)  &  1.000103(42)  &   0      \\
  11 &   0.708394 &  0.708360(35)  &  0.999952(49)  &   1      \\
  12 &   0.695433 &  0.695412(37)  &  0.999970(53)  &   1      \\
  13 &   0.683390 &  0.683478(26)  &  1.000129(38)  &   0      \\
  14 &   0.683390 &  0.683430(46)  &  1.000060(67)  &   1      \\
  15 &   0.695433 &  0.695342(43)  &  0.999870(62)  &   0      \\
  16 &   0.708394 &  0.708390(40)  &  0.999994(57)  &   1      \\
  17 &   0.695433 &  0.695401(44)  &  0.999955(63)  &   1      \\
  18 &   0.683390 &  0.683318(42)  &  0.999895(61)  &   0      \\
  19 &   0.683390 &  0.683426(31)  &  1.000053(46)  &   0      \\
  20 &   0.695433 &  0.695408(51)  &  0.999965(73)  &   1      \\
  21 &   0.768360 &  0.768363(30)  &  1.000005(39)  &   1      \\
  22 &   0.722100 &  0.722037(36)  &  0.999912(49)  &   0      \\
  23 &   0.695433 &  0.695449(44)  &  1.000024(64)  &   1      \\
  24 &   0.695433 &  0.695465(43)  &  1.000046(61)  &   1      \\
  25 &   0.722100 &  0.722112(44)  &  1.000017(60)  &   1      \\
\hline
\end{tabular}}
\label{table:gii}
\end{table*}

\subsection{Lattices}
In our main study, the loop model (\ref{Zloop}) is regularized on a
two-dimensional honeycomb lattice.  As remarked before, the coordination
number of the honeycomb lattice is three and allows a unique
decomposition of closed graphs into an ensemble of polygons.  We
constructed the honeycomb lattice from its dual, i.e., hexagonal, or
triangular, lattice.  The latter, which, in contrast to the former, is a
Bravais lattice, is spanned by two vectors of equal length, making an
angle of 60\textdegree.  We have chosen the lattice spacing of the dual
lattice to be unity, $a_\triangle=1$.  The lattice spacing of the
honeycomb lattice is then fixed to be $a_{\varhexagon} =
1/\sqrt{3}=0.5773\ldots$.  Euclidean distances on the honeycomb lattice
are measured in units of $a_{\varhexagon}$.  The number of lattice sites
on the dual lattice, i.e., the volume, is taken to be $V_\triangle =
L^2$, where $L$ denotes the number of lattice sites in any of the two
independent directions on the hexagonal lattice.  This parameter $L$
features as the linear lattice size variable in all our further
considerations, including our finite-size scaling analyses.  Under the
dual construction, the volume of the honeycomb lattice picks up a factor
of two, so that $V_{\varhexagon} = 2 V_\triangle$, while the number $B$
of links is unchanged, $B_\triangle = B_{\varhexagon} =3 V$.

\begin{figure}
\centering
%\psfrag{x}[t][t][2][0]{$M$}
%\psfrag{y}[t][t][2][0]{PDF$(M)$}
\includegraphics[angle=-90,width=0.8\textwidth]{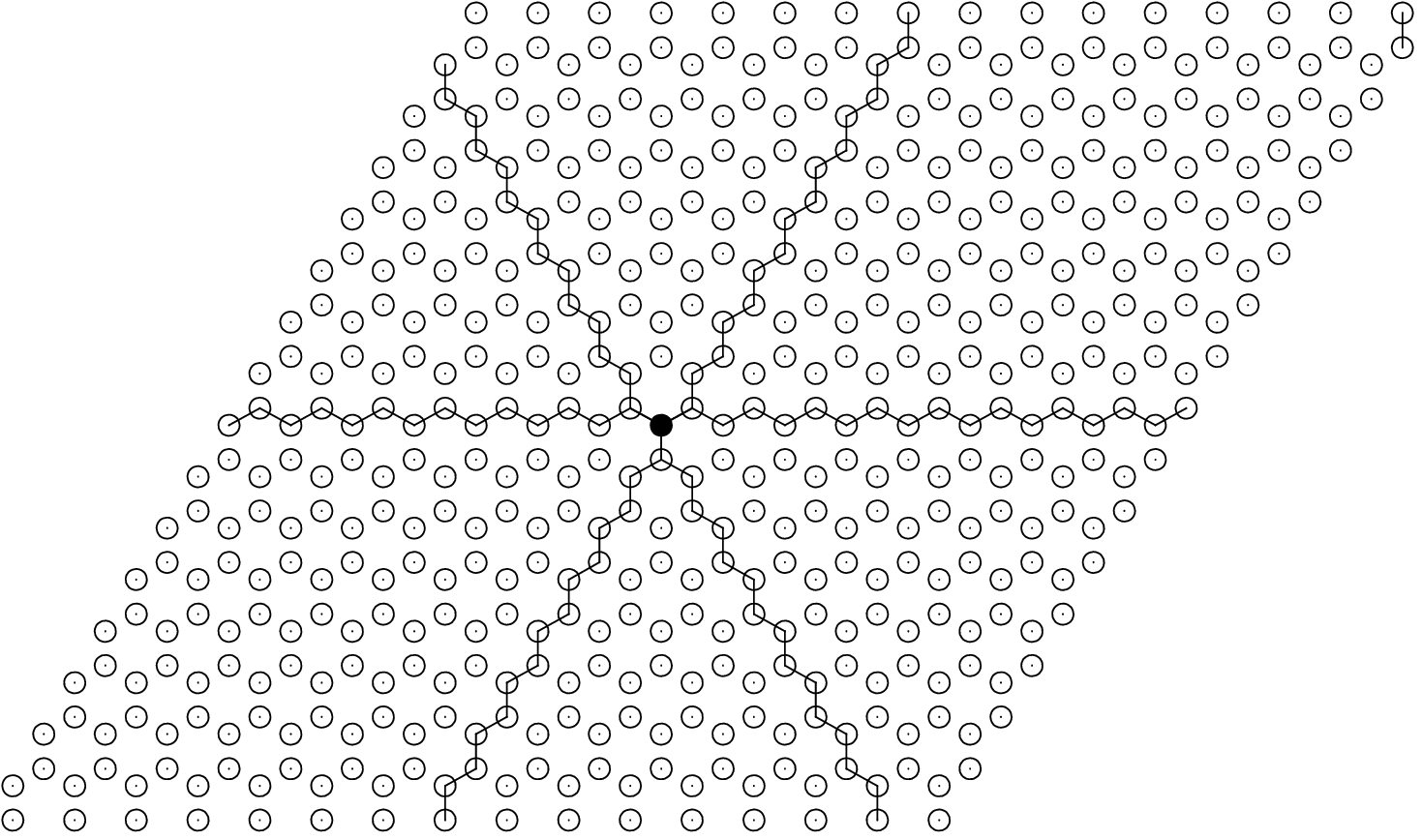}
\caption{Compact honeycomb lattice with periodicity in three directions.
  The three operations mapping the site $x_i$ marked by a full circle
  onto itself through shift operations are indicated by the three
  polygons, each winding the lattice once.  Note the set bond in the upper right corner, 
which by the periodicity of the lattice belongs to the polygon winding the lattice in the northwest direction.
\label{fig:honeycomb}}
\end{figure}

We constructed a honeycomb lattice that is compact with periodicity of
$2L$ in three directions.  Let $\Delta_\mu(x_i)$ denote the shift
operation that connects a site $x_i$ to its nearest neighbor in the
$\mu=1,2,3$ direction.  Then there exist three product operations, each
involving $2L$ such shifts, that form the identity and map any site
$x_i$ onto itself, see Fig.~\ref{fig:honeycomb} for an example with
$L=16$.

\subsection{Observables}
In our Monte Carlo simulations, we analyzed the two sectors of the
model, i.e., configurations with and without an open chain, separately.
We implemented a search algorithm that uniquely decomposes
(disconnected) closed graphs into polygons along the links of the
underlying honeycomb lattice.  Each polygon is assigned a length
parameter $n$, denoting the number of sites visited.  While tracing the
loops, we also record for each site $x_{i_k}$ visited by a polygon, the
direction to the next site $x_{i_{k+1}}$ in that polygon.  This makes
possible to determine whether a polygon winds the (periodic) lattice in
any of the three possible directions.  Specifically, we determine for
each polygon its winding number, which is a topological invariant,
telling how often it winds the (periodic) lattice in a given direction.
Note that because the endpoints of the worm erase or draw bonds, the
worm algorithm can change the winding number of a configuration.  A
nonzero winding number is the signal for loop percolation.  As in
percolation theory, such ``infinite'' polygons are usually excluded
from measurements of variables not connected to percolation observables
to facilitate finite-size scaling analyses.

\begin{figure}
\centering
\psfrag{x}[t][t][2][0]{$n$}
\psfrag{y}[t][t][2][0]{$\ell_n$}
\includegraphics[angle=-90,width=0.8\textwidth]{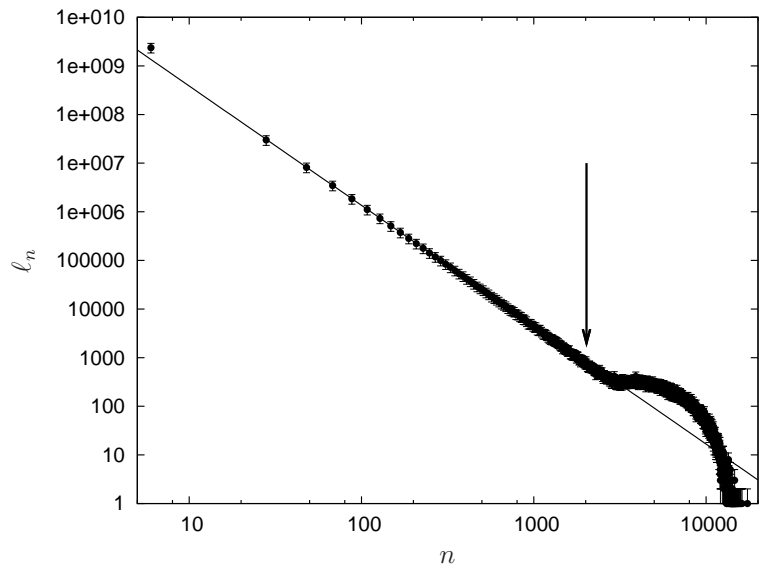}
\caption{Log-log plot of the loop distribution $\ell_n$ as a function of
  the loop length $n$ on the honeycomb lattice of size $L=352$ at the
  critical point.  The arrow indicates the minimal length $n_0=2022$
  from Table~\ref{table:min}.  The straight line proportional to
  $n^{-2/D-1}$ with $D=\frac{11}{8}$ is put through the data points by
  hand to show the expected behavior (\ref{elln}).
  \label{fig:loop_distribution}}
\end{figure}

For observables analyzed on lattices of fixed size, an even stringent
upper bound on chain lengths is required.  To this end, we monitored
during the Monte Carlo runs the length $n$ of winding loops.  The
minimum loop length found in the time series, each involving $10^7$
sweeps of the lattice, represents a natural upper bound on (open and
also closed) chain lengths to be included in such analyses.  The minimum
values $n_0$ at the critical point for the honeycomb lattice of several
sizes are given in Table~\ref{table:min}.
\begin{table*}
  \caption{Length $n_0$ of the shortest loop winding the honeycomb lattice of
    linear extent $L$ recorded during $10^7$
    sweeps of the lattice at the critical point.}
\label{table:min}
\centerline{
\begin{tabular}{c|rrrrrrrrrrr}
\hline $L$ & 32 & 64 & 96 & 128 & 160 & 192 & 224 & 256 &
288 & 320 & 352 \\
$n_0$ & 78 & 196 & 346 & 510 & 690 & 854 & 1138 & 1320 & 1602 & 1820 & 2022 \\
\hline
\end{tabular}}
\end{table*}
The length $n_0(L)$ can be equally well interpreted as the length of the
\textit{largest} loop that can be realized within a lattice of linear
size $L$.  Stated differently, $n_0(L)$ indicates the loop length up to
which the scaling law (\ref{elln}) applies, see
Fig~\ref{fig:loop_distribution}.  Scaling implies that this length
increases with the linear lattice size as $n_0(L) \sim L^D$.  The
results in Table~\ref{table:min} satisfy this scaling with
$D=\frac{11}{8}$, as anticipated, see Table~\ref{table:On}.

We measure the loop distribution $\ell_n$ by compiling a histogram of
loop lengths during long Monte Carlo runs
\begin{equation}
\ell_n = \frac{1}{\mathcal{N}} \sum_m \delta_{n_m,n},
\end{equation}
where $m$ enumerates the polygons measured with $\sum_m = \mathcal{N}$
denoting the total number of polygons measured, and $n_m$ is the length
of the $m$th polygon.  Figure~\ref{fig:loop_distribution} shows the
results of such measurements at the critical point on the largest
lattice considered, i.e., $L=352$.
Loops at all scales are observed.  The bump at the end of the
distribution followed by a rapid falloff is typical for such
distributions measured on a finite lattice with periodic boundary
conditions.  Finally, we determine for each polygon its center of mass
(\ref{center}) as well as the square radius of gyration (\ref{Rg}).

We next turn to the analysis of configurations containing an open chain
in addition to polygons.  Unlike configurations without one, those with
an open chain cannot always be uniquely decomposed, even on a honeycomb
lattice.  Consider, for example, an open chain with an endpoint housing
three bonds.  It is then not clear whether it represents a single,
self-intersecting chain, or a chain and a separate polygon which touch
each other at the chain endpoint.  The decomposition is unique on a
honeycomb lattice only if both chain endpoints contain just one bond.
To minimize the arbitrariness associated with tracing out non-unique
open chains, we omitted from the measurements chains with both endpoints
housing three bonds.  This only concerns a small fraction of all chain
configurations measured of about two percent at the critical point.
When only one endpoint of the open chain contains three bonds, we
interpret the configuration as representing an open chain and a separate
polygon.  This is similar in spirit to the closing of open SAWs, see
the argument leading to Eq.~(\ref{ellz}) and Fig.~\ref{fig:saw}. As for
polygons, we determined the square radius of gyration $R_\mathrm{g}^2$
of open chains as a function of the chain length $n$, as well as their
square end-to-end distance $R_\mathrm{e}^2$ also as a function of $n$.

\subsection{Details of simulation}
The simulations are carried out on honeycomb lattices of linear extent
ranging from $L=32$ to $L=352$ in steps of $\Delta L=32$.  Each
simulation consists of $10^7$ sweeps, where a single sweep is defined as
$V_{\varhexagon} = 2 V_\triangle = 2L^2$ local bond updates.  An
additional 10\% of the sweeps is used for thermalization. The
accumulated computer time used for the simulations amounts to a few
weeks on a single workstation.

\section{Simulation results}
\label{sec:results}
To check our code, we make use of the celebrated Kramers-Wannier duality
for the two-dimensional Ising model \cite{KrWa}.  This duality asserts
that the loop gas, or HT, representation of the O($N=1$) model on a
two-dimensional lattice at the same time represents the model in the
standard spin representation on the dual lattice.  To picture a spin
configuration on the dual lattice, imagine drawing bonds between any
pair of nearest neighbor spins on that lattice which are in the same
spin state.  When no bond exists between two nearest neighbor spins,
they are in different spin states.  A HT bond in a given loop
configuration can be interpreted as indicating a broken bond between the
two nearest neighbor spins living on the dual lattice, on either side of
the HT bond on the original lattice.  That is, a HT bond indicates that
the two corresponding spins on the dual lattice are in different spin
states.  Given this transcription, a loop can then be pictured as
forming the boundary of a cluster of nearest neighbor spins on the dual
lattice which are all in the same spin state.  This implies that loop
configurations containing just a few bonds correspond to ordered spin
configurations on the dual lattice.  More generally, under the dual map,
the high-temperature phase of the loop model, in which, for sufficiently
high temperatures, only a few small loops are present, maps onto the
low-temperature phase of the spin model, where, for sufficiently low
temperatures, large spin clusters can be found.  The two temperatures
can be related by noting that a HT bond carries a factor $K = \tanh
(\beta)$, while a nearest neighbor pair on the dual lattice of unlike
spins on each side of the HT bond carries a Boltzmann weight
$\exp(-2\tilde \beta)$, so that \cite{KrWa}
\begin{equation} 
\label{bb}
K  = {\rm e}^{-2\tilde \beta} .
\end{equation} 
Note that the dual map as described here is special to two
dimensions and cannot be generalized to higher dimensions.

Not any loop configuration can be resolved in a spin configuration on
the dual lattice.  Loop configurations containing, for example, a single
loop winding the lattice once do not, given the periodic boundary
conditions, translate into a spin configuration on the dual lattice.  It
is straightforward to see that when the winding number of a given loop
configuration is even in any of the three directions, such a
transcription is possible up to a factor Z(2) which is chosen at random.

\begin{figure}
\centering
\psfrag{x}[t][t][2][0]{$M$}
\psfrag{y}[t][t][2][0]{PDF$(M)$}
\includegraphics[angle=-90,width=0.8\textwidth]{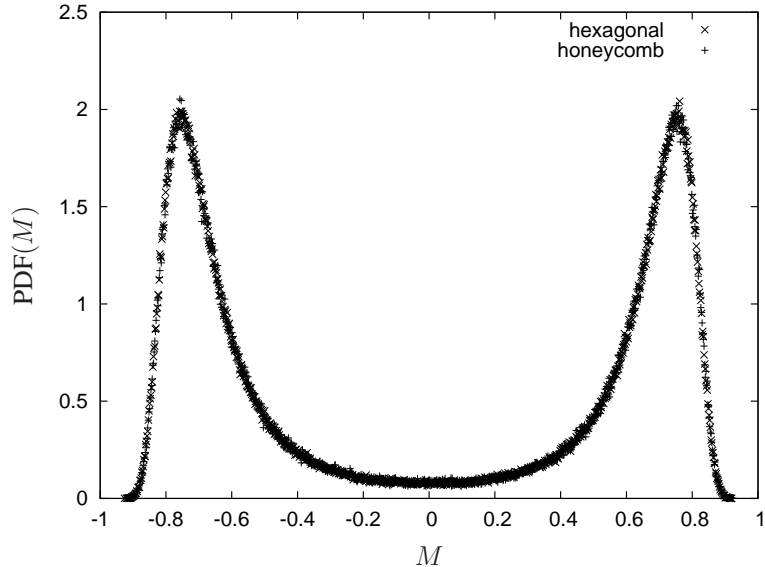}
\caption{Probability distribution function (PDF) of the magnetization
  (M) measured at $K_\mathrm{c}$, i.e., at the critical
    point of the infinite lattice, on lattices of size $L=32$.  One
  data set, marked by $\times$, is obtained using the Swendsen-Wang
  cluster algorithm on the hexagonal lattice, the other data set, marked
  by $+$, is obtained using the worm algorithm on the honeycomb lattice,
  and then transcribed to the hexagonal lattice.  Both sets are seen to
  nicely blend, as expected by duality.
  \label{fig:magdis}}
\end{figure}

To demonstrate duality and also the correctness of our Monte Carlo
simulations, we measured the magnetization $M$ of the Ising model at the
critical point, using the two representations.  For the standard spin
representation of the Ising model on the hexagonal lattice, we use the
Swendsen-Wang cluster update \cite{SwendsenWang}, while for the loop
model, or HT representation, on the honeycomb lattice, we
use the worm algorithm. Only those loop configurations are considered
that can be mapped onto a spin configuration on the hexagonal lattice.
Figure~\ref{fig:magdis} attests that, as expected, the two distinct data
sets nicely merge.
\begin{figure}
\centering
\psfrag{x}[t][t][2][0]{$\tilde{\beta}$}
\psfrag{y}[t][t][2][0]{$\langle |M| \rangle$}
\includegraphics[angle=-90,width=0.8\textwidth]{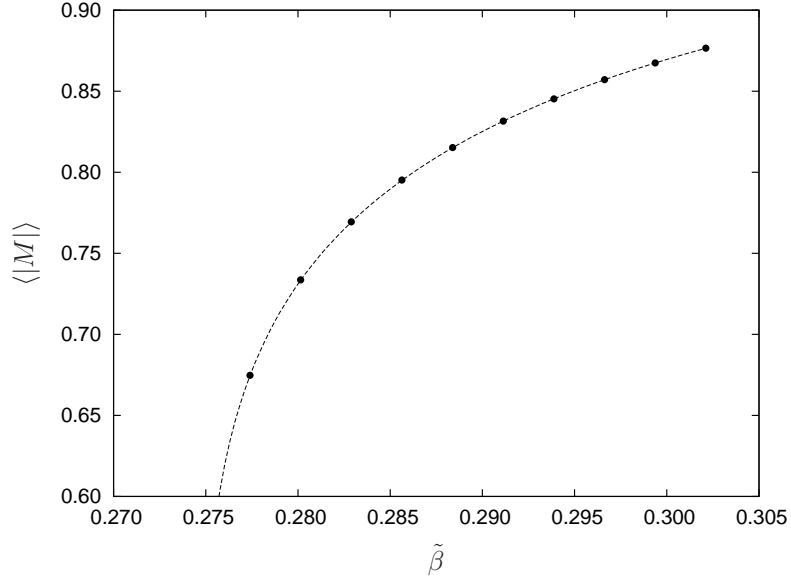}
\caption{Average absolute value of the magnetization on a hexagonal
  lattice as a function of the inverse temperature $\tilde{\beta}$.  The curve
  gives the exact result (\ref{potts}) due to Potts.
  \label{fig:magnetization}}
\end{figure}

As a further illustration, we display in Fig.~\ref{fig:magnetization}
the average of the absolute value of the magnetization $| M |$ as
obtained from the dual map as a function of the inverse temperature
$\tilde{\beta}$ of the Ising model on the hexagonal lattice introduced
in Eq.~(\ref{bb}).  The simulation itself was carried out on the $L=96$
honeycomb lattice for inverse dual temperatures $\tilde{\beta} >
\tilde{\beta}_\mathrm{c}$, where by Eqs.~(\ref{bb}) and (\ref{Kc})
\begin{equation} 
\tilde{\beta}_\mathrm{c} \equiv \tfrac{1}{4} \ln 3 = 0.27465 \ldots \, .
\end{equation} 
The plotted curve corresponds to the exact calculation by Potts
\cite{Potts}
\begin{equation}
\label{potts}
M^8(\tilde{\beta})=1- \frac{16 \mathrm{e}^{ -12 \tilde{\beta}}}{\left(1+3 \mathrm{e}^{ -4 \tilde{\beta}
  }\right)\left(1-\mathrm{e}^{ -4 \tilde{\beta}}\right)^3}
\end{equation}
for the functional form of the magnetization of the Ising model as a
function of the inverse temperature $\tilde{\beta}$ on an infinite
hexagonal lattice.  The agreement between the theoretical curve and the
data is seen to be excellent.

\begin{figure}
\centering \psfrag{x}[t][t][2][0]{$L$} 
\psfrag{y}[t][t][2][0]{$U_L$}
\includegraphics[angle=-90,width=0.8\textwidth]{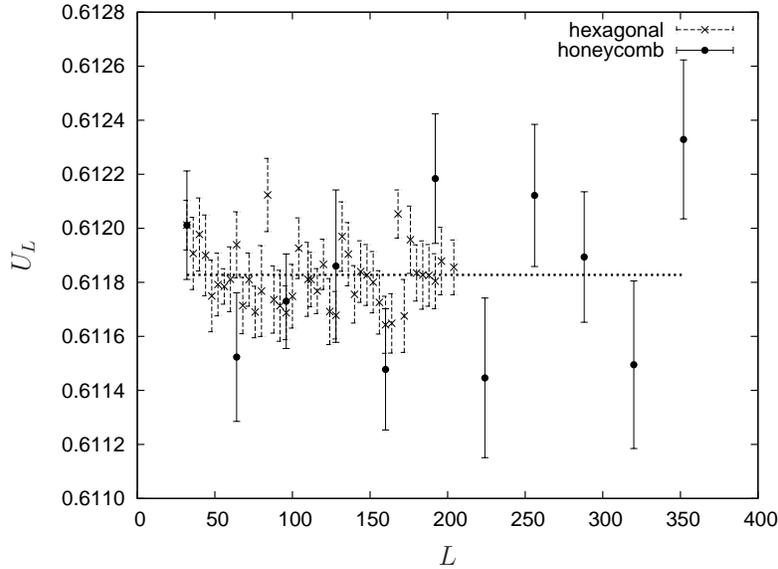}
\caption{Binder parameter $U_L$ as a function of the linear lattice size
  $L$.  One data set is obtained using the Swendsen-Wang cluster
  algorithm on the hexagonal lattice, the other set is obtained using
  the worm algorithm on the honeycomb lattice, and then transcribed to the hexagonal lattice.  The straight line
  indicates the high-precision result obtained in
  Ref.~\protect\cite{KB} by transfer matrix methods.
  \label{fig:binder_plot}}
\end{figure}

As a final check on the correctness of our Monte Carlo simulations, we
measured the Binder parameter
\begin{equation}
U_L \equiv 1- \frac{1}{3} \frac{\langle M^4 \rangle}{\langle M^2 \rangle^2} ,
\end{equation}
which involves the second and fourth moments of the magnetization $M$,
see Fig.~\ref{fig:binder_plot}.  The definition of this parameter is
such that for a Gaussian theory it vanishes.  The critical value of the
Binder parameter of the Ising model on a hexagonal lattice with periodic
boundary conditions is known extremely well from transfer matrix
calculations \cite{KB}, \textit{viz.} $U_L=0.61182773(1)$.  Having no
scaling dimension, $U_L$ does not change with lattice size in leading
order.  One-parameter fits to the data lead to the estimates $U_L =
0.611822 (73)$ with $\chi^2/\textsc{dof} = 1.45$ for the loop gas on
the honeycomb lattice (transcribed to the hexagonal lattice), and $U_L =
0.611815 (17)$ with $\chi^2/\textsc{dof} = 1.01$ for the standard spin
representation on the hexagonal lattice.  Both estimates, based on
different representations of the Ising model and obtained using
different update algorithms, are in excellent agreement with the
high-precision result.

\begin{figure}
\centering
\psfrag{x}[t][t][2][0]{$\tilde{\beta}$}
\psfrag{y}[t][t][2][0]{$I_L$}
\includegraphics[angle=-90,width=0.8\textwidth]{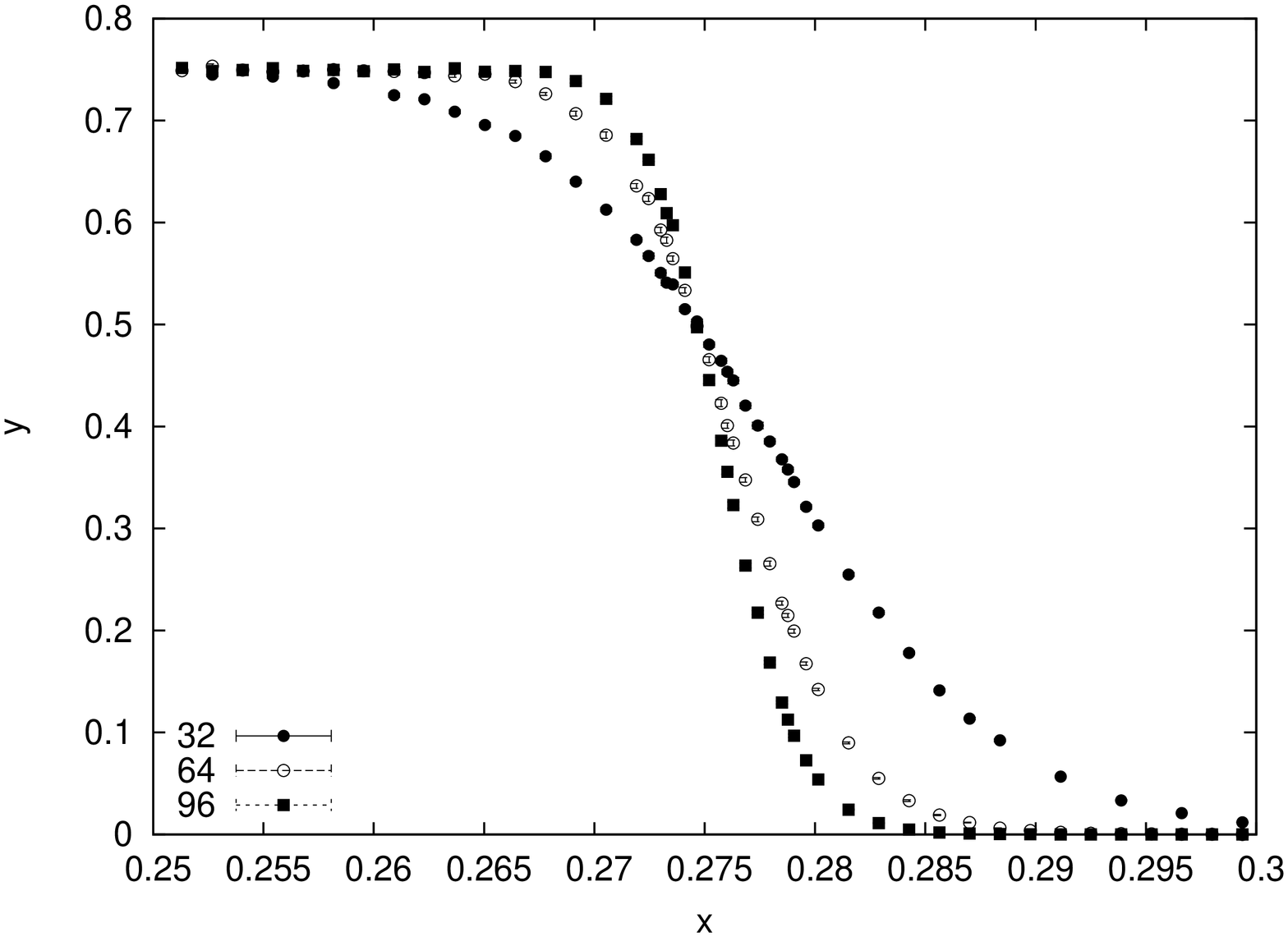} \\
\psfrag{x}[t][t][2][0]{$(\tilde{\beta}-\tilde{\beta}_\mathrm{c}) L^{1/\nu}$}
\psfrag{y}[t][t][2][0]{$I_L$}
\includegraphics[angle=-90,width=0.8\textwidth]{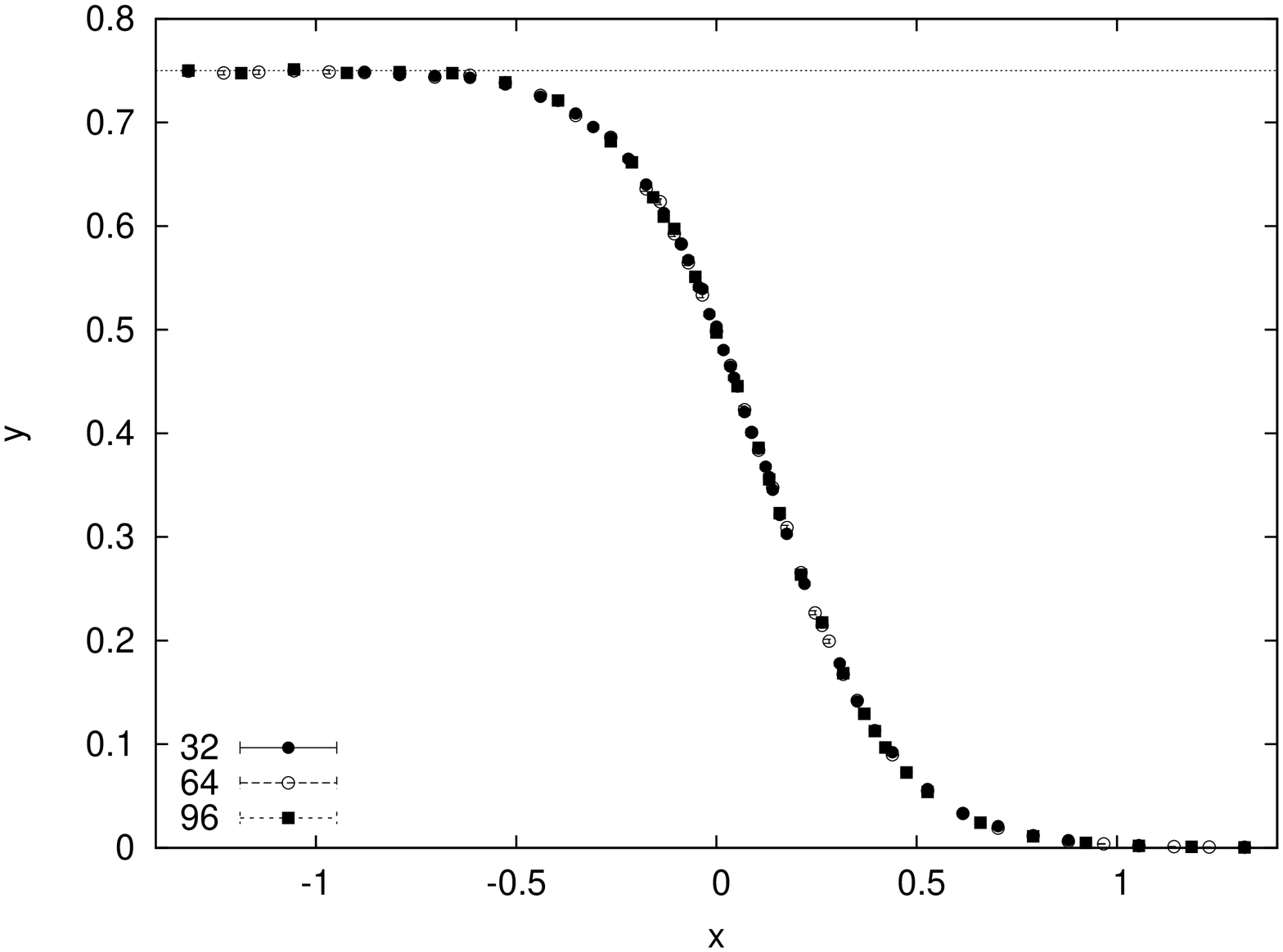}
\caption{\textit{Top panel}: Average $I_L(\tilde{\beta})$ as a function of the
  inverse dual temperature $\tilde{\beta}$ an lattices of linear extent $L=32,64,96$.
  \textit{Bottom panel}: Same data reploted as a function of the scaling
  variable $(\tilde{\beta} - \tilde{\beta}_\mathrm{c}) L^{1/\nu}$ with $\nu =1$.  In the
  limit $\tilde{\beta} \to 0$, $I_L(\tilde{\beta})$ tends to $\frac{3}{4}$ (straight line).
  \label{fig:operc}}
\end{figure}

We next turn to estimators of physical observables which exploit the
nature of the worm update algorithm and which can be naturally measured
in this scheme.  As first estimator we introduce the binary variable
that records whether a loop configuration can be mapped onto a spin
configuration on the dual lattice, or not.  If a map exists, this
observable is assigned the value zero, else it is assigned the value
unity.  The top panel in Fig.~\ref{fig:operc} shows the average
$I_L(\tilde{\beta})$ of this observable as a function of the inverse
dual temperature $\tilde{\beta}$ introduced in Eq.~(\ref{bb}) on lattices
of linear extent $L=32,64,96$.
For large $\tilde{\beta}$, where mostly only a few small loops are
present, loop configurations can typically be mapped onto spin
configurations on the dual lattice and $I_L(\tilde{\beta})$ is small,
tending to zero in the limit $\tilde{\beta} \to \infty$.  When
$\tilde{\beta}$ is lowered, larger loops appear and eventually loops can
be found that wind around the lattice.  As mentioned above, when, for
example, a single loop does so once, a transcription is impossible, and
the value unity is recorded in the time series of measurements.  This
explains the increase of $I_L(\tilde{\beta})$ with decreasing
$\tilde{\beta}$.  In the limit $\tilde{\beta} \to 0$, where loops are
abundant, this observable is seen to tend to an asymptotic value (very
close to) $\frac{3}{4}$.  In words, the ratio of the number of loop
configurations with an odd winding number to those with an even winding
number tends to $\frac{3}{4}$ in the zero-temperature limit, where it is
recalled that, although impossible for loop configurations with an odd
winding number, configurations with an even winding number can be mapped
onto a spin configuration on the dual lattice.  The observable
$I_L(\tilde{\beta})$ has no scaling dimension and plays a role similar
to the Binder parameter.  Finite-size scaling implies that it depends
not on the inverse temperature $\tilde{\beta}$ and the lattice size $L$
independently, but only on the combination $(\tilde{\beta}
-\tilde{\beta}_\mathrm{c}) L^{1/\nu}$, with $\nu$ the correlation length
exponent.  The bottom panel in Fig.~\ref{fig:operc}, which displays the
same data, but now as a function of this scaling variable with $\nu=1$,
shows that finite-size scaling is satisfied.

\begin{figure}
\centering
\psfrag{x}[t][t][2][0]{$L$}
\psfrag{y}[t][t][2][0]{$I_L'(\tilde{\beta}_\mathrm{c})$}
\includegraphics[angle=-90,width=0.8\textwidth]{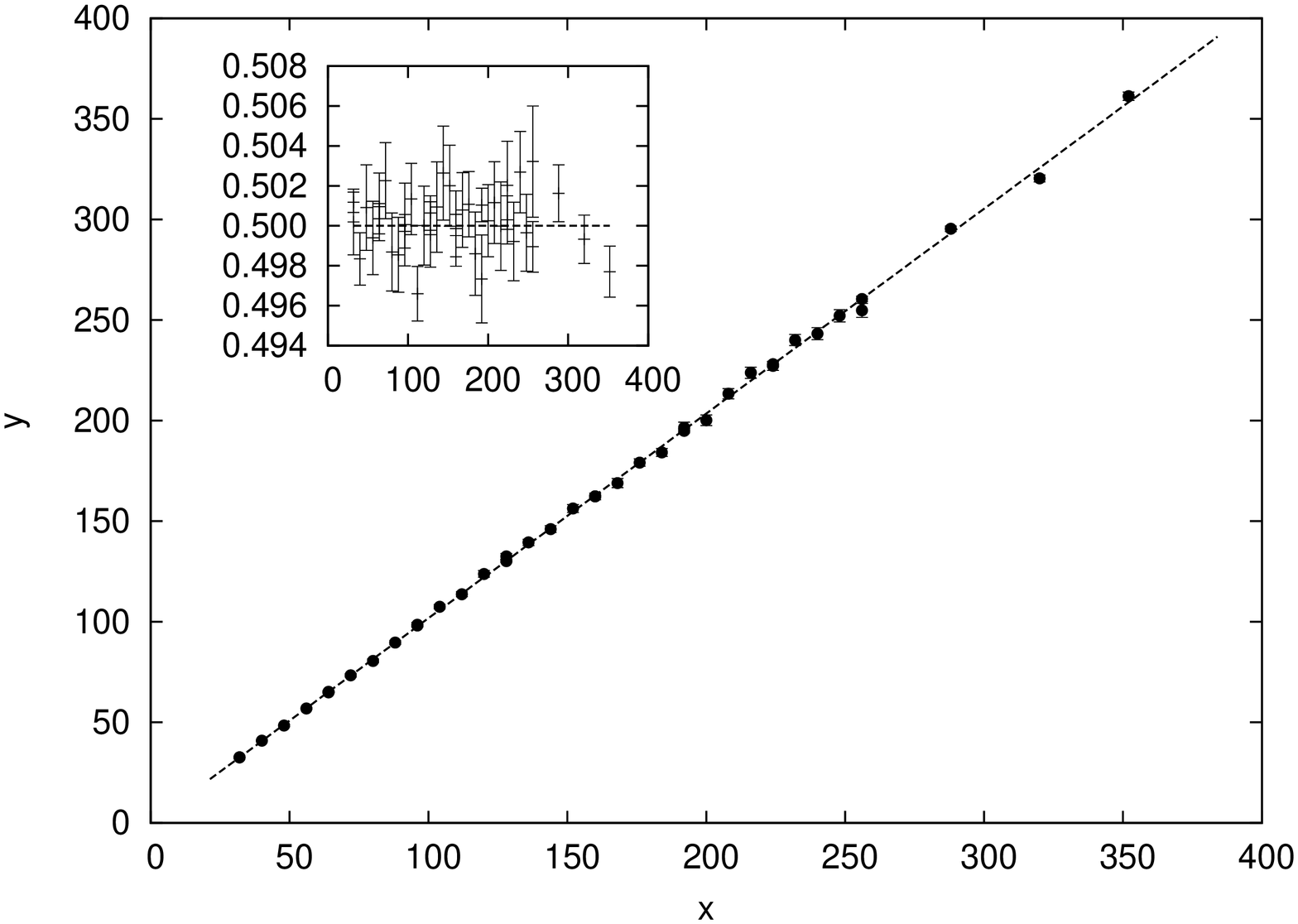}
\caption{\textit{Inset}: Binary observable $I_L(\tilde{\beta}_\mathrm{c})$ as a
  function of the linear lattice size $L$.  \textit{Main panel}:
  Derivative $I'_L(\tilde{\beta}_\mathrm{c})$ of $I_L(\tilde{\beta})$ with respect to
  $\tilde{\beta}$ evaluated at the critical point $\tilde{\beta}_\mathrm{c}$ as a
  function of $L$.
  \label{fig:operc_derivative}}
\end{figure}

To quantify this statement, we measured $I_L(\tilde{\beta})$ at the
critical point on lattices of different size, see inset of
Fig.~\ref{fig:operc_derivative}.  A one-parameter fit to the data leads
to the estimate $I_L(\tilde{\beta}_\mathrm{c})=0.50024(21)$ with
$\chi^2/\textsc{dof} = 0.851$.  That is, half of the critical
configurations have an odd winding number.  Since $I_L(\tilde{\beta})$
depends only on the scaling variable $(\tilde{\beta} -
\tilde{\beta}_\mathrm{c}) L^{1/\nu}$, differentiation of
$I_L(\tilde{\beta})$ with respect to $\tilde{\beta}$ allows estimating
$1/\nu$, see Fig.~\ref{fig:operc_derivative}.  A two-parameter fit to
the data gives as estimate for the slope $1/\nu = 1.0001 (15)$ with
$\chi^2/\textsc{dof} = 1.01$, in excellent agreement with the expected
value $\nu =1$.

\begin{figure}
\centering
\psfrag{x}[t][t][2][0]{$n$}
\psfrag{y}[t][t][2][0]{$z_n$}
\includegraphics[angle=-90,width=0.8\textwidth]{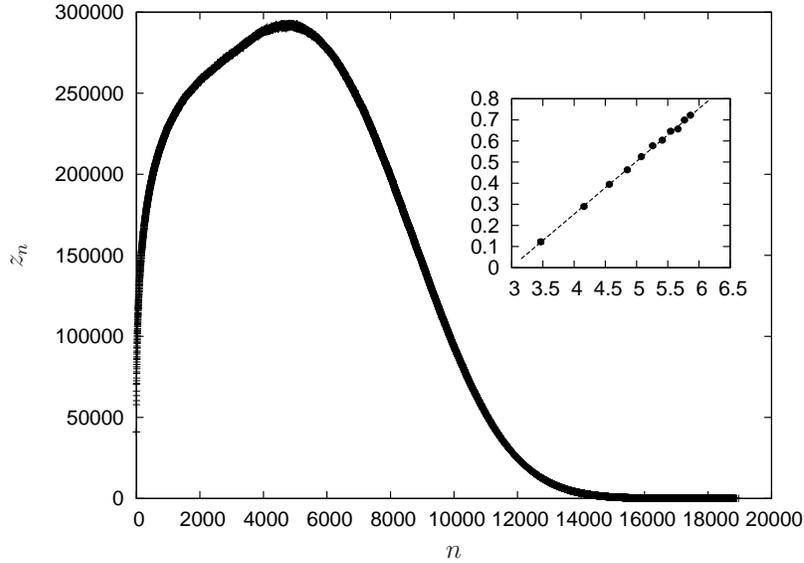}
\caption{Distribution $z_n$ of open chains of arbitrary end-to-end
  distance at $K_\mathrm{c}$ on a lattice of linear size $L=352$ as a
  function of chain length $n$.  \textit{Inset}: Log-log plot of the
  inverse of the integrated observable $\sum_n z_n = \chi$ divided by the
  volume $L^2$, i.e., of $L^2/\chi$, measured on lattices of
  different size, as a function of $L$.
  \label{fig:poldis}}
\end{figure}

As mentioned above, the great virtue of the worm algorithm is that it
also generates open chains.  Figure~\ref{fig:poldis} shows the
distribution $z_n$, introduced in Eq.~(\ref{zn_def}), of chains of $n$
steps with arbitrary end-to-end distance $r$ measured at the critical
point on a lattice of linear size $L=352$.  By Eq.~(\ref{Gsum}) and the
definition of the magnetic susceptibility $\chi$ in terms of the
correlation function given below Eq.~(\ref{xi2}), the sum $\sum_n z_n$
gives $\chi$ at the critical point.  According to finite-size scaling,
$\chi(K_\mathrm{c})$ scales with lattice size $L$ as
\begin{equation} 
\chi (K_\mathrm{c}) = \sum_n z_n \sim L^{\gamma/\nu}.
\end{equation} 
The inset of Fig.~\ref{fig:poldis} shows a log-log plot of the ratio
$L^2/\chi$ as a function of the lattice size.  A linear two-parameter
fit to the data gives the estimate $2- \gamma/\nu = 0.2498 (26)$ with
$\chi^2/\textsc{dof} = 1.87$, in agreement with the exact result $\eta
= 2- \gamma/\nu = \frac{1}{4}$.

\begin{figure}
  \centering \psfrag{x}[t][t][2][0]{$n$} \psfrag{y}[t][t][2][0]{$\frac{1}{5}\langle R^2_\mathrm{e}
    \rangle, \langle
    R_\mathrm{g}^2\rangle$}
  \includegraphics[angle=-90,width=0.8\textwidth]{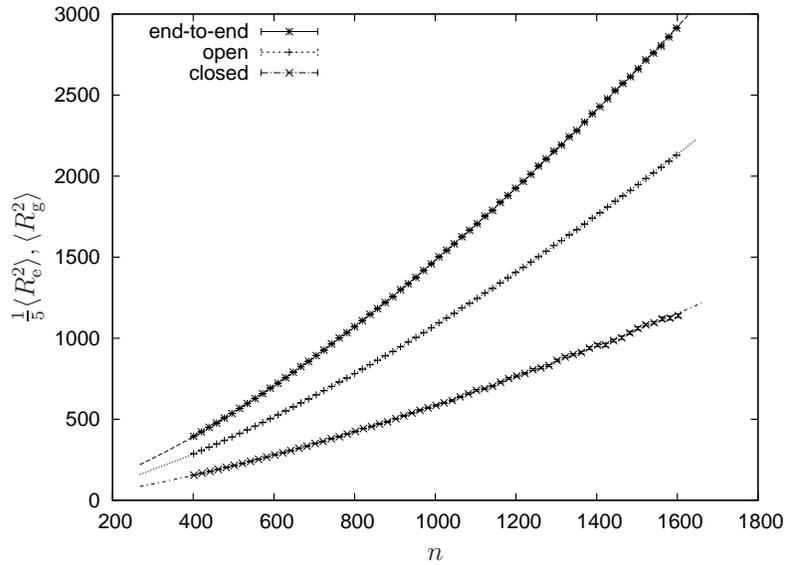}
\caption{Average square end-to-end distance $\langle R^2_\mathrm{e} \rangle$
  (scaled by a factor of five for readability) of open chains, as well as the
average square radius of gyration $\langle
  R_\mathrm{g}^2\rangle$ of open and closed  chains, 
all shown
  as a function of their length $n$ and measured at the critical point
  on a lattice of linear size $L=352$.
  \label{fig:rms2_etc_scaling}}
\end{figure}

An important characteristic of the chains generated by the worm
algorithm, whether closed or open, is their Hausdorff, or fractal,
dimension $D$.  Figure~\ref{fig:rms2_etc_scaling} shows the average
square radius of gyration $\langle R_\mathrm{g}^2\rangle$ of closed and
open chains as a function of $n$, as well as the average square
end-to-end distance $\langle R_\mathrm{e}^2\rangle$ of open chains, also
as a function of $n$.  In obtaining an accurate estimate of the fractal
dimension from these and corresponding data measured on lattices of
different size, we face two restrictions.  The first is that the scaling
(\ref{rms}) only holds asymptotically, i.e., for sufficiently large $n$.
To lift this restriction somewhat, we include the leading correction to
scaling in Eq.~(\ref{rms}) by writing
\begin{equation} 
\label{rsm_fit}
\langle R_\mathrm{g,e}^2\rangle = a n^{2/D} \left(1 + \frac{b}{n} 
\right) .
\end{equation} 
As for SAWs \cite{0409355}, we expect this leading correction term to be
analytic and inversely proportional to $n$.  The length of a chain is,
on the other side, bounded by the number of lattice sites available on a
finite lattice.  Figure~\ref{fig:poldis} shows that, as a result, even
at the critical point, the number $z_n$ of open chains of arbitrary end-to-end
distance falls exponentially for large $n$ on a finite lattice.  To
obtain an estimate for the infinite lattice, chains that exceed a
maximum length $n_0$ where they start to notice the finite extent of the
lattice must be ignored.  As measure of this maximum, we take the length
of the shortest polygon winding the lattice recorded in the time series,
see Table~\ref{table:min}.  To be on the safe side, the maximum value of
$n$ included in the data considered for fitting is chosen to be about
$0.8 \,  n_0$, while the minimum value it taken to be $n_\mathrm{min}
= 100$.
\begin{table*}
\caption{Estimates of the fractal dimension $D$ obtained through
  three-parameter fits to the average square radius of gyration $\langle
  R_\mathrm{g}^2\rangle$ of open chains (left) and to the average
  square end-to-end distance $\langle R_\mathrm{e}^2\rangle$ (right)
  obtained on lattices of size $L$.}
\label{table:rms}
\centerline{
\begin{tabular}{l|lll|lll}
\hline
$L$ & $D$ & $D/D_\mathrm{exact}$ & $\chi^2/\textsc{dof}$  & $D$ & $D/D_\mathrm{exact}$ & $\chi^2/\textsc{dof}$        \\
\hline
   160   &      1.485(84)  &    1.080(61)   & 0.424  & 1.43(14)   &    1.04(10)   & 0.682     \\ 
   192   &      1.4213(62) &    1.033(19)   & 0.493  & 1.475(46)  &    1.073(33)  & 0.588     \\ 
   224   &      1.3822(96) &    1.0052(70)  & 0.301  & 1.381(16)  &    1.004(11)  & 0.288     \\ 
   256   &      1.3672(67) &    0.9943(49)  & 0.499  & 1.364(12)  &    0.9924(90) & 0.453     \\ 
   288   &      1.3762(44) &    1.0008(32)  & 0.867  & 1.3733(79) &    0.9987(58) & 0.624     \\ 
   320   &      1.3738(34) &    0.9991(24)  & 0.850  & 1.3733(67) &    0.9988(49) & 0.791     \\ 
   352   &      1.3755(31) &    1.0004(23)  & 0.697  & 1.3789(55) &    1.0028(40) & 0.985     \\ 
\hline                                                                                            
$\infty$ &      1.3747(19) &    0.9998(14)  &          & 1.3752(35) &    1.0002(25) &            \\  
\hline
\end{tabular}}
\end{table*}
Table~\ref{table:rms} summarizes the estimates for the fractal dimension
$D$ obtained through three-parameter fits using the form
(\ref{rsm_fit}).  The last line in the table gives the weighted average
of the estimates for $L > 200$, which are all consistent with a
constant, i.e., $L$-independent value.  The final results are in
excellent agreement with the exact value $D_\mathrm{exact}=\frac{11}{8}$
predicted by Saleur and Duplantier \cite{SD}.  Drawing on numerical work
by Cambier and Nauenberg \cite{CaNau}, Vanderzande and Stella
\cite{VdzandeStellaJP} provided early indirect support for this
prediction.  Direct numerical support was first provided by Dotsenko
{\it et al.}~\cite{Dotsenkoetal}, and more recently in
Ref.~\cite{geoPotts} using other, percolationlike estimators and a
plaquette update.

\begin{figure}
\centering
\psfrag{x}[t][t][2][0]{$\ln(L)$}
\psfrag{y}[t][t][2][0]{$\langle n_\mathrm{w} \rangle$}
\includegraphics[angle=-90,width=0.8\textwidth]{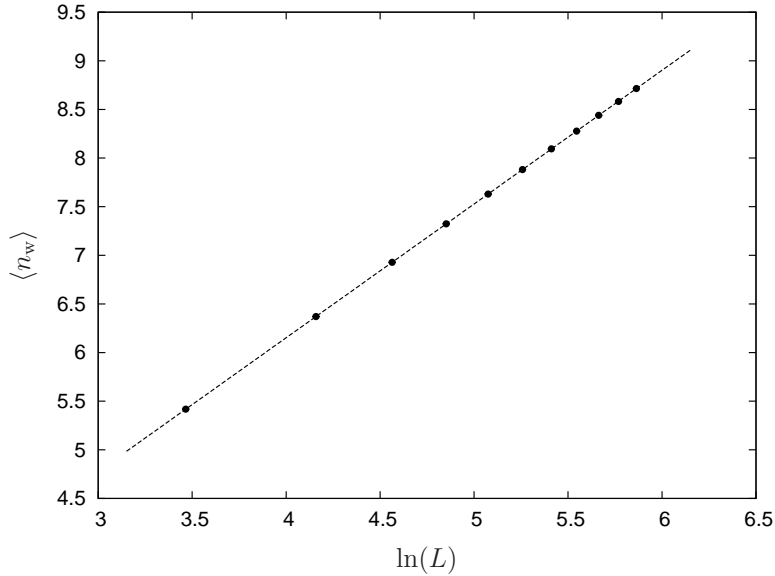}
\caption{Log-log plot of the average length $\langle n_\mathrm{w} \rangle$ of loops winding the honeycomb lattice of linear extent $L$.
  \label{fig:log_mean_length_of_percolated_loop_versus_length}}
\end{figure}

Polygons that wind the lattice have been excluded from the above
measurements.  However, as in percolation theory, the fractal dimension
of polygons (clusters) at the critical point can also be determined by
exclusively focussing on such winding polygons (percolating clusters).
Because these polygons are long, no corrections to scaling as in
Eq.~(\ref{rsm_fit}) need to be included.
Figure~\ref{fig:log_mean_length_of_percolated_loop_versus_length} shows
the average length $\langle n_\mathrm{w} \rangle$ of loops winding the
honeycomb lattice as a function of lattice size $L$.
A linear fit to the data obtained on lattices of size $L=32$ up to
$L=352$ using $\langle n_\mathrm{w} \rangle \propto L^D$ yields $D=1.37504(32)$
with $\chi^2/\textsc{dof}=1.13$, in excellent agreement with the
predicted result.  Note the extra digit of precision achieved here in
comparison to the above estimates.

\begin{figure}
\centering
\psfrag{x}[t][t][2][0]{$L$}
\psfrag{y}[t][t][2][0]{$\Pi_L$}
\includegraphics[angle=-90,width=0.8\textwidth]{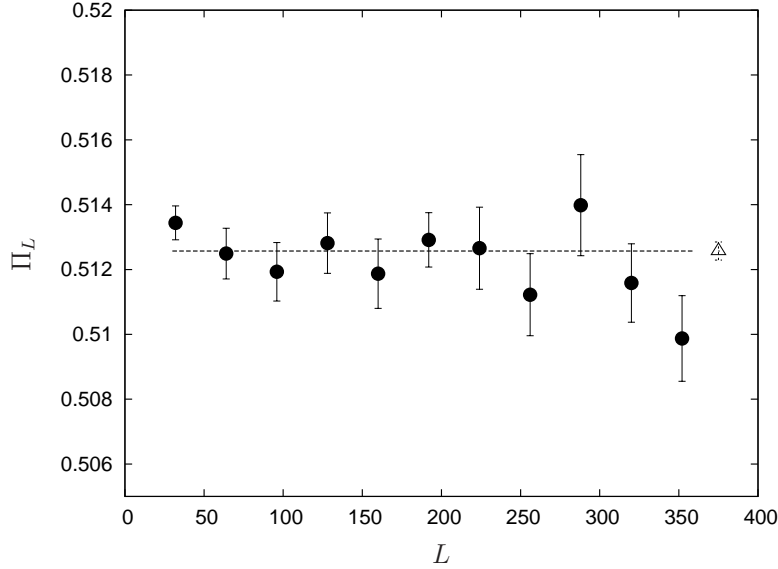}
\caption{Probability $\Pi_L$
that one or more loops wind the honeycomb lattice of linear size $L$ at the critical point.
The most right symbol in the figure denotes the weighted average with error bars. 
\label{fig:percolation_threshold}}
\end{figure}

In Fig.~\ref{fig:percolation_threshold}, we show the probability $\Pi_L$
that one or more loops wind the lattice of size $L$ at the critical
point.  The data show no finite-size effects.  A one-parameter fit to
the data with $L$ ranging from $L=32$ up to $L=352$ gives $\Pi_L =
0.51257(27)$ with $\chi^2/\textsc{dof}=1.08$.  This probability is
slightly larger than the probability
$I_L(\tilde{\beta}_\mathrm{c})=0.50024(21)$ of finding a configuration
with odd winding number because $\Pi_L$ also includes configurations
with (nonzero) even winding number.

To detail this further, we give in Table~\ref{table:winding} the
probability $P_w$ of finding a configuration with winding number $w$
measured on lattices of sizes $L=32$ and $L=160$ at the critical point
$K=K_\mathrm{c}$ and also in the low-temperature phase at $K =
K_\mathrm{LT}$ with ($N=1$)
\begin{equation}
\label{KLT}
K_\mathrm{LT} \equiv \left[2 - (2 -N)^{1/2}\right]^{-1/2}  ,
\end{equation} 
defining the low-temperature branch of the O($N$) model on a honeycomb
lattice \cite{Nienhuis}.  At this temperature, the bond fugacity becomes
unity for the Ising model so that each configuration carries the same
weight according to the partition function (\ref{Zloop}).  The loop
model thus reduces to a purely geometric random model where a bond
update is always accepted.  By Eq.~(\ref{bb}), this temperature
corresponds to vanishing inverse dual temperature $\tilde{\beta}=0$ of
the Ising model on the dual lattice where spins are oriented up or down
at random.  In this limit, the dual model becomes equivalent to random
site percolation at the percolation threshold $p_\mathrm{c} =
\frac{1}{2}$, and the polygons on the honeycomb lattice denote the
boundaries of the occupied sites on the hexagonal lattice \cite{SD}.
\begin{table*}
  \caption{Probability of finding a closed loop configuration with winding number $w$ on honeycomb lattices of size $L=32$ and $L=160$ at the critical point $K_\mathrm{c}$ and at $K_\mathrm{LT}$ where $\tilde{\beta}=0$. }
\label{table:winding}
\centerline{
\begin{tabular}{l|ll|ll}
\hline
& \multicolumn{2}{c|}{$K_\mathrm{c}$} & \multicolumn{2}{c}{$K_\mathrm{LT}$} \\
\hline
$w$ & $P_w(L=32)$ & $P_w(L=160)$ & $P_w(L=32)$ & $P_w(L=160)$ \\
\hline
0 & 0.48747(89) & 0.48802(74) & 0.15764(29)  & 0.15836(29) \\
 1 & 0.50029(88) & 0.49960(73) & 0.74579(34)  & 0.74545(53) \\
 2 & 0.01221(12) & 0.01236(14) & 0.09214(16)  & 0.09171(36) \\
 3 & 0.0000307(70) & 0.0000299(64) & 0.004369(70)  & 0.004413(76) \\
 4 &        0        &      0          & 0.0000588(64)  & 0.000060(11) \\
\hline
\end{tabular}}
\end{table*}
For the two lattice sizes considered, we have not recorded any
configuration with winding number $w \geq 5$.  For the critical theory,
we did not even observe a single configuration with $w=4$.  The data
show no finite-size effects.  The sum of the measured probabilities
$P_w$ with $w$ odd equals $0.50032(89)$ for $L =32$ and $0.49963(73)$
for $L = 160$ in the case of the critical theory, and $0.75016(41)$ for
$L =32$ and $0.74987(60)$ for $L =160$ at $\tilde{\beta}=0$.  Since
these numerical results are perfectly consistent with the fractions
$\frac{1}{2}$ and $\frac{3}{4}$, it is tempting to speculate that these
are in fact exact results.

\begin{figure}
\centering
\psfrag{x}[t][t][2][0]{$r$}
\psfrag{y}[t][t][2][0]{$10^5 P_n(r)$}
\includegraphics[angle=-90,width=0.8\textwidth]{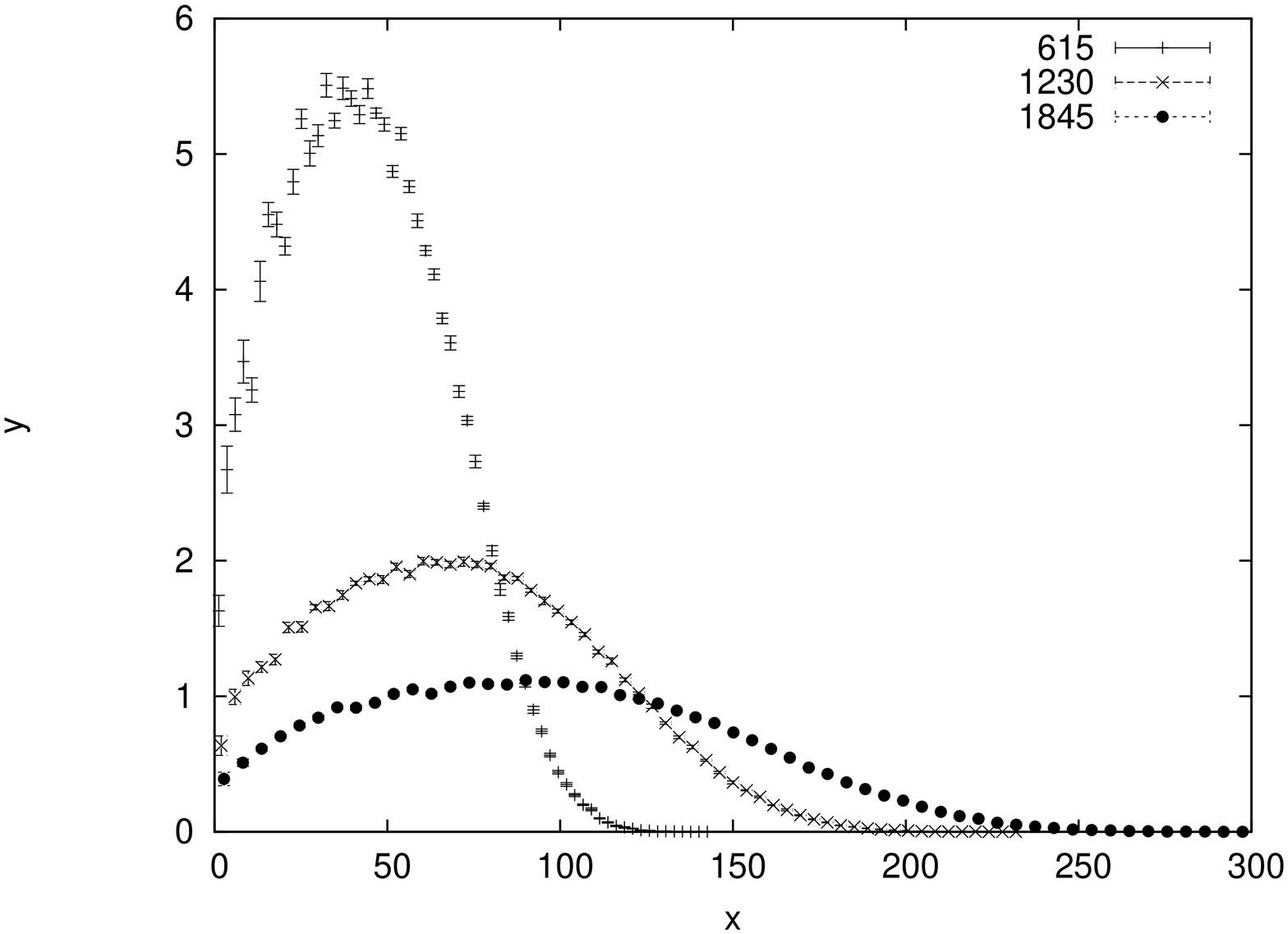}
\\
\psfrag{x}[t][t][2][0]{$r/n^{1/D}$}
\psfrag{y}[t][t][2][0]{$n^{2/D} P_n$}
\includegraphics[angle=-90,width=0.8\textwidth]{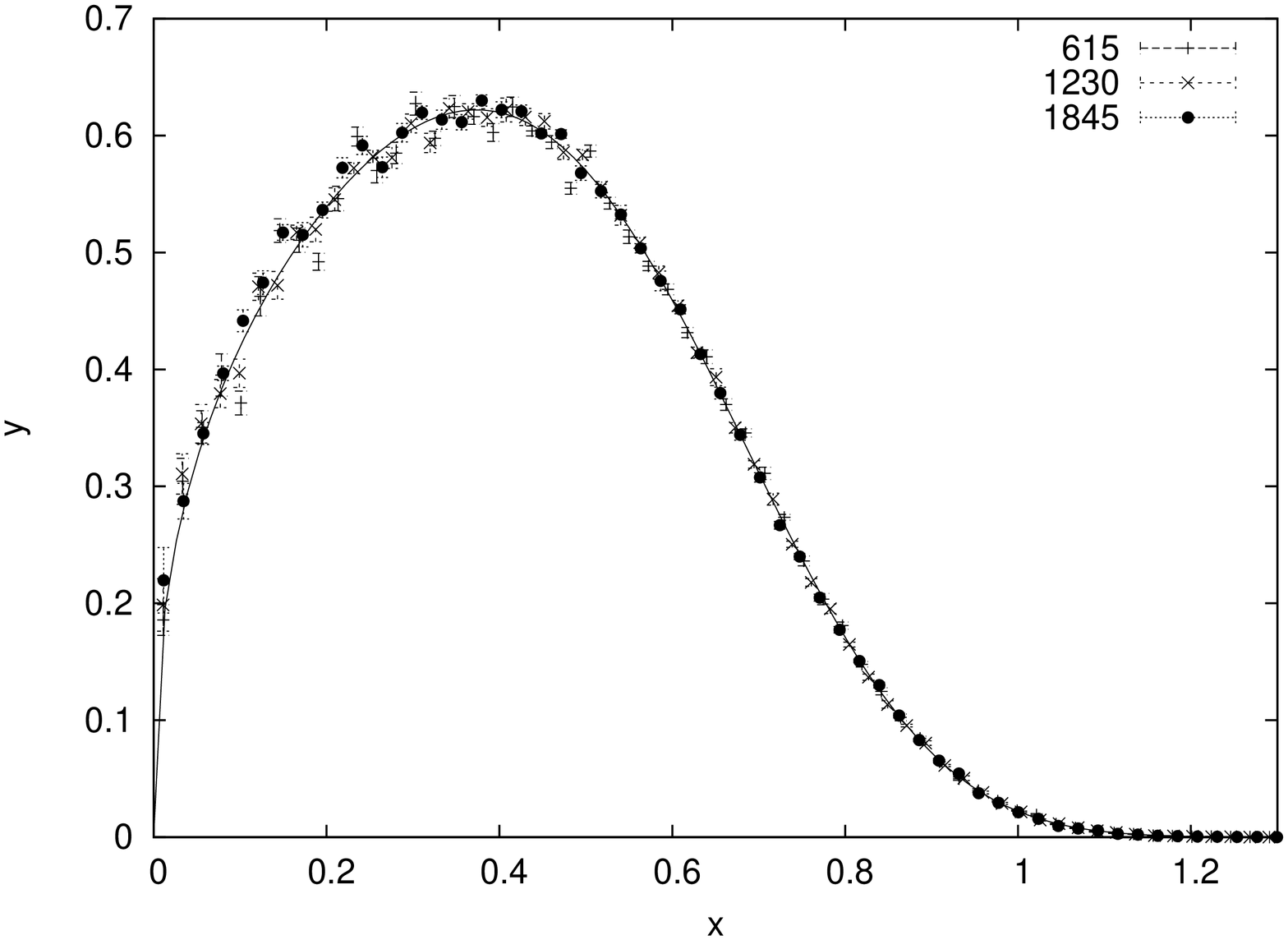}
\caption{\textit{Top panel}: Distribution $P_n(r)$ (multiplied by a factor of $10^5$ 
for convenience) as a function of the
  end-to-end distance $r$ for chains of length $n=615, 1230, 1845$
  measured on the largest lattice considered, \textit{viz.}
  $L=352$. \textit{Bottom panel}: Rescaled data shown in top panel.  The
  curve through the data points is based on a three-parameter fit to the
  data using the predicted form (\ref{universal}).
  \label{fig:end_to_end_distribution_functions}}
\end{figure}

As final observable, we consider the distribution function $P_n(r)$ of
the end-to-end distance $r$ introduced in Eq.~(\ref{P}).  The top panel
in Fig.~\ref{fig:end_to_end_distribution_functions} shows this
distribution as a function of $r$ for chains of length $n=615, 1230$, and
$1845$ measured on the largest lattice considered, \textit{viz.} $L=352$.
Each of the distributions are normalized according to Eq.~(\ref{norm})
with $d=2$.  If our finite-size scaling conjecture (\ref{P}) holds for
the O(1) loop gas, the data for various $n$ should collapse onto a
universal curve when $n^{d/D} P_n$ with $d=2$ and $D=\frac{11}{8}$ is
plotted as a function of the scaling variable $r/n^{1/D}$.  This is
indeed what we observe, see bottom panel in
Fig.~\ref{fig:end_to_end_distribution_functions}.  The analyses of data
measured on smaller lattices, typically with three equidistant chain
lengths, give similar results.  For SAWs it has been suggested
\cite{Cloizeaux_book} that the whole scaling function can be
approximated by a single function
\begin{equation} 
\label{universal}
\mathcal{P}(t) = a t^\vartheta \exp \left(-b t^\delta\right) ,
\end{equation} 
with parameters $a$ and $b$.  This scaling function slightly generalizes
the form originally proposed by Fisher \cite{Fisher}, where $b=1$ was
assumed.  The exponent $\delta$ is related to the fractal dimension $D$
of the SAWs by the Fisher law \cite{Fisher}
\begin{equation} 
\delta = \frac{1}{1-1/D} .
\end{equation} 
We have cast this law in a form that allows generalization to arbitrary
critical O($N$) loop gases, where it is recalled that for SAWs, which
are described by the O($N \to 0$) model, $\nu=1/\sigma D$ with
$\sigma=1$, but for arbitrary $-2 \leq N \leq 2$, $\sigma\neq 1$.  With
$\delta$ given the predicted value $\delta= \frac{11}{3}$ for $N=1$, a
three-parameter fit to the data yields a surprisingly good approximation
of the entire scaling function, see Table~\ref{table:3fit}.  The
resulting estimate for $\vartheta$ is in excellent agreement with the
predicted value $\vartheta = \frac{3}{8}$.  Note that the normalization
(\ref{norm}) translates into the normalization
\begin{equation} 
1 = \Omega_d \int_0^\infty \mathrm{d} t \, t^{d-1}  \mathcal{P}_n(t) 
\end{equation} 
of the scaling function.  With the explicit form (\ref{universal}), this
gives a relation between the three parameters $a,b$, and $\vartheta$.
Our estimates for these parameters satisfy this relation within
statistical errors, so that the fits effectively involve only two free
parameters.
\begin{table*}
  \caption{Results of three-parameter fits to the
    data in Fig.~\ref{fig:end_to_end_distribution_functions} using the predicted 
    form (\ref{universal}) with $\delta= \frac{11}{3}$. The last line in the table 
    gives the weighted average of
    the estimates for $L > 200$, which are all consistent with a constant,
    i.e., $L$-independent value.}
\centerline{
\begin{tabular}{l|lllll}
\hline
$L$ & $a$ & $b$ & $\vartheta$ & $\vartheta/\vartheta_\mathrm{exact}$ & $\chi^2/\textsc{dof}$         \\
\hline
 160        &  1.017(08)   &  3.819(18)  &  0.3970(73) &  1.058(19)  &  1.80  \\
 192        &  0.992(11)   &  3.798(24)  &  0.377(10)  &  1.006(27)  &  2.74  \\
 224        &  1.0070(94)  &  3.820(19)  &  0.3835(85) &  1.022(22)  &  2.31  \\
 256        &  1.0019(59)  &  3.815(12)  &  0.3793(56) &  1.011(15)  &  1.48  \\
 288        &  1.0022(50)  &  3.807(10)  &  0.3799(47) &  1.013(12)  &  1.35  \\
 320        &  0.9950(47)  &  3.790(10)  &  0.3733(44) &  0.995(11)  &  1.35  \\
 352        &  0.9973(51)  &  3.814(10)  &  0.3721(46) &  0.992(12)  &  1.53  \\
\hline
 $\infty$   &  0.9994(26)  &  3.8096(55) &  0.3769(24) &  1.0050(64) &      \\
\hline
\end{tabular}}
\label{table:3fit}
\end{table*}

\section{Outlook and discussion}
\label{sec:outlook}
The worm algorithm may potentially turn the loop gas approach to
fluctuating fields on a lattice into a viable alternative for
numerically simulating lattice field theories.  Being an alternative,
the algorithm calls for new estimators of physical observables, a few of
which we have described here.  Up to now, the loop gas approach has been
widely adopted only in the de~Gennes $N \to 0$ limit of the O($N$)
model, which describes self-avoiding random walks.  As demonstrated in
this paper, concepts developed to describe such random walks can be
generalized to arbitrary O($N$) models.  A next step in advancing the
worm algorithm, which we hope to explore in a future publication, is to
include gauge fields.  Finally, although field theories that contain
fermions can also be expanded in a strong-coupling series, and are in
principle amenable to the worm algorithm, it is an open question whether
the algorithm can deal with the sign problem.  A first step towards this
problem, to which we also hope to return in a future publication, is to
simulate the critical O($N$) model for $N=-1$ on a honeycomb lattice, so
that according to the HT representation (\ref{Zloop}) each polygon
caries a minus sign.  That is, in this particular representation, the
O($N=-1$) model exhibits the sign problem in its pristine form.

\section*{Acknowledgement} 
Work supported in part by the Deutsche Forschungsgemeinschaft (DFG)
under grant No.\ JA483/23-2 and the EU RTN-Network `ENRAGE':
``Random Geometry and Random Matrices: From Quantum Gravity to
Econophysics'' under grant No.~MRTN-CT-2004-005616.


\begin{thebibliography}{99}
\bibitem{Feynman} R. P. Feynman, Rev. Mod. Phys. {\bf 20}, 367 (1948).
\bibitem{Stanley} H. E. Stanley, \textit{Introduction to Phase
    Transitions and Critical Phenomena} (Oxford University Press, New
  York, 1971). 
\bibitem{BB} B. Berg and D. Foerster, Phys. Lett. B \textbf{106}, 323 (1981).
\bibitem{DaHa} C. Dasgupta and B. I. Halperin, Phys. Rev. Lett.
  \textbf{47}, 1556 (1981). 
\bibitem{HM} W. Helfrich and W. M\"uller, in {\it Continuum Models of
    Discrete Systems} (University of Waterloo Press, Waterloo, Ontario,
  Canada, 1980), p. 753.  See also F. Rys and W. Helfrich, J. Phys. A:
  Math. Gen. \textbf{15}, 599 (1982).
\bibitem{Karowski_etal} M. Karowski, H. J. Thun, W. Helfrich, and F. S. Rys,
J. Phys. A: Math. Gen. \textbf{16}, 4073 (1983); M. Karowski and F. S. Rys,
J. Phys. A: Math. Gen. \textbf{19}, 2599 (1986). 
\bibitem{HJK} T. Hofs\"ass, W. Janke, and H. Kleinert, Phys. Lett.
  A \textbf{105}, 463 (1984); W. Janke and H. Kleinert, Phys. Lett.
  A \textbf{128}, 463 (1988). 
\bibitem{Elser} V. Elser, \textit{Topics in Statistical Mechanics},
Ph.D. Thesis, University of California, Berkeley (1984).
\bibitem{ProkofevSvistunov} N. Prokof'ev and B. Svistunov,
Phys. Rev. Lett. \textbf{87}, 160601 (2001).
\bibitem{geoPotts} W. Janke and A. M. J. Schakel, Nucl. Phys. B [FS]
  {\bf 700}, 385 (2004).
\bibitem{ht} W. Janke and A. M. J. Schakel, Phys. Rev. Lett.
  \textbf{95}, 135702 (2005).
\bibitem{StauferAharony} D. Stauffer and A. Aharony, {\it Introduction
    to Percolation Theory}, 2nd edition (Taylor \& Francis, London,
  1994).
\bibitem{Nienhuis_rev} B. Nienhuis, in: {\it Phase Transitions and
    Critical Phenomena}, edited by C. Domb and J. L. Lebowitz (Academic,
  London, 1987), Vol. 11, p. 1.
\bibitem{deGennes} P. G. de Gennes, Phys. Lett. A \textbf{38}, 339
  (1972).
\bibitem{Fisher} M. E. Fisher, J. Chem. Phys. \textbf{44}, 616 (1966).
\bibitem{Nienhuis} B. Nienhuis, Phys. Rev. Lett. {\bf 49}, 1062 (1982).
\bibitem{Vanderzande} C. Vanderzande, J. Phys. A: Math. Gen. {\bf 25},
  L75 (1992).
\bibitem{Jensen04} I. Jensen, J. Phys. A: Math. Gen. {\bf 37}, 5503
  (2004).
\bibitem{Cloizeaux} J. des Cloizeaux, Phys. Rev. A {\bf 10}, 1665
  (1974).
\bibitem{deGennesbook}  P. G. de Gennes, \textit{Scaling Concepts in Polymer Physics}
  (Cornell University Press, Ithaca, 1979).
\bibitem{McKenzieMoore} D. S. McKenzie and M.A. Moore, J. Phys. A: Math.
  Gen.  \textbf{4}, L82 (1971).
\bibitem{ProkofevSvistunov_com} N. Prokof'ev and B. Svistunov,
Phys. Rev. Lett. \textbf{96}, 219701 (2006).
\bibitem{JanSmit} J. Smit, \textit{Introduction to Quantum Fields on a
Lattice} (Cambridge University Press, Cambridge, 2002).
\bibitem{DMNS} E. Domany, D. Mukamel, B. Nienhuis, and A. Schwimmer,
  Nucl.  Phys. B \textbf{190}, 279 (1981).
\bibitem{CPS} L. Chayes, L. P. Pryadko, and K. Shtengel, Nucl. Phys. B
\textbf{570}, 590 (2000).
\bibitem{Gabriel} C. Gabriel, {\it Dynamical properties of the worm
    algorithm}, {\it Diplomarbeit}, Technische Universit\"at Graz
  (2002).
\bibitem{Wolff} U. Wolff, Nucl. Phys. B \textbf{810}, 491 (2009).
\bibitem{KrWa} H. A. Kramers and G. H. Wannier, Phys. Rev. {\bf 60}, 252 (1941).
\bibitem{SwendsenWang} R. H. Swendsen and J. S. Wang, Phys. Rev. Lett.
  {\bf 58}, 86 (1987).
\bibitem{Potts} R. B. Potts, Phys. Rev. \textbf{88}, 352 (1952). 
\bibitem{KB} G. Kamieniarz and H. W. J. Bl\"ote, J. Phys. A: Math. Gen.
  \textbf{26}, 201 (1993).
\bibitem{0409355} S. Caracciolo, A. J. Guttmann, I. Jensen, A.
  Pelissetto, A. N. Rogers, and A. D. Sokal, J. Stat. Phys.  \textbf{120},
  1037 (2005).
\bibitem{SD}  H. Saleur and B. Duplantier, Phys. Rev. Lett. {\bf 58}, 2325
(1987).
\bibitem{CaNau} J. L. Cambier and M. Nauenberg, Phys. Rev. B {\bf 34},
8071 (1986).
\bibitem{VdzandeStellaJP} C. Vanderzande and A. L. Stella, J. Phys. A:
  Math. Gen.  {\bf 22}, L445 (1989).
\bibitem{Dotsenkoetal} V. S. Dotsenko, M. Picco, P. Windey, G. Harris,
E. Martinec, and E. Marinari, Nucl. Phys. B {\bf 448} [FS], 577 (1995).
\bibitem{Cloizeaux_book} J. des Cloizeaux and G. Jannink,
  \textit{Polymers in Solution: Their Modelling and Structure} (Oxford
  University Press, Oxford, 1989).
\end{thebibliography}
\end{document}